\documentclass[10pt]{article}
\usepackage{graphicx}
\usepackage{epstopdf}
\makeatletter
\newcommand\appendix@section[1]{%
  \refstepcounter{section}%
  \orig@section*{Appendix \@Alph\c@section: #1}%
  \addcontentsline{toc}{section}{Appendix \@Alph\c@section: #1}%
}
\let\orig@section\section
\g@addto@macro\appendix{\let\section\appendix@section}
\makeatother
\usepackage{amsmath}
\usepackage{mathtools}
\usepackage{latexsym}
\usepackage{morefloats}
\usepackage[numbers, square, comma, sort&compress]{natbib}
\advance\oddsidemargin by -\textwidth
\advance\oddsidemargin by -1in
\evensidemargin=\oddsidemargin
\def\fsl#1{\setbox0=\hbox{$#1$}           
   \dimen0=\wd0                                 
   \setbox1=\hbox{/} \dimen1=\wd1               
   \ifdim\dimen0>\dimen1                        
      \rlap{\hbox to \dimen0{\hfil/\hfil}}      
      #1                                        
   \else                                        
      \rlap{\hbox to \dimen1{\hfil$#1$\hfil}}   
      /                                         
   \fi}                                         %
\makeatletter
\def\@maketitle{\newpage
 \null
 {\normalsize \tt \begin{flushright} 
  \begin{tabular}[t]{l} 
  \end{tabular}
 \end{flushright}}
 \begin{center}
 \vskip 2em
 {\LARGE \@title \par} \vskip 2.5em {\large \lineskip .5em
 \begin{tabular}[t]{c}\@author 
 \end{tabular}\par} 
 \end{center}
} 

\newcommand{\vereq}[2]{\lower3pt\vbox{\baselineskip1.5pt \lineskip1.5pt
\ialign{$\m@th#1\hfill##\hfil$\crcr#2\crcr\sim\crcr}}}

\newcommand\eqsecnum{
\@newctr{equation}[section]
\renewcommand\theequation{\arabic{section}.\arabic{equation}}%
}

\newbox\tempboxa
\newdimen\captionboxsubcount 
\def\capsize#1{\captionboxsubcount=#1pt}
\newdimen\captionboxsub
\captionboxsub=\hsize \advance\captionboxsub by -\captionboxsubcount
\advance\captionboxsub by -\captionboxsubcount
\long\def\@makecaption#1#2{
 \setbox\@tempboxa\hbox{\footnotesize #1: #2}
 \ifdim \wd\@tempboxa >\captionboxsub 
\rightskip=\captionboxsubcount \leftskip=\captionboxsubcount 
  \footnotesize #1: #2 
\else \hbox to\hsize{\hfil\box\@tempboxa\hfil} 
 \fi}
\makeatother
\eqsecnum
\capsize{30}

\title{{\Large\bf 
Light-Front Dynamic Analysis of Bound States in Scalar Field Model
 \vspace{5mm}} }
\author{ Chueng-Ryong Ji and Yukihisa Tokunaga\\
{\small{\it Department of Physics, North Carolina State University, Raleigh, NC 27695, USA}}}

\begin{document}

\maketitle

\begin{abstract}
The light-front dynamics (LFD) of the scalar field model theory is analyzed to solve the two-body bound-state problem. The light-front two-body bound-state equation is extended to the full LFD kernel including the ladder, cross-ladder, stretched-box, and particle-antiparticle creation/annihilation effects to study the contributions of higher Fock-states. The light-front two-body equation is also modified by the term corresponding to the self-energy corrections and counter-terms. Using the variational principle, we obtain the numerical result of the binding energy B versus the coupling constant $\alpha$ for various mass ratios of the constituent particles including the cases of non-zero exchange particle mass. We also discuss the correlation between the mass spectrum and the corresponding bound-state wavefunction. 
\end{abstract}

\section{Introduction}
For an accurate calculation of the spectra and wavefunctions of hadrons, it is important to include the fundamental relativistic effects such as the correct relativistic energy-momentum relation, retardation effects, and particle-antiparticle creation and annihilation. Since the quarks and gluons inside the hadrons have negligible masses and interact very strongly among themselves via quantum chromodynamics (QCD), the relativistic effects play a significant role in analyzing the mass spectra and the wavefunction related observables (e.g. form factors, generalized parton distributions, etc.), in particular, for the low-lying hadrons. Although the ultimate goal is to analyze the bound-state problem in QCD, prior to getting into the real complicated nature of QCD, we may first investigate much simpler bound-state system provided by the Wick-Cutkosky model \cite{Cutkosky:1954ru,Wick:1954eu} and the similar scalar field model theories. 

C. Savkli et al. \cite{Savkli:2004bi,Savkli:1999ui,Savkli:2002fj} have already used a powerful numerical approach known as the Feynman-Schwinger representation approach and investigated the scalar field model theory which they called $\chi^2\phi$-theory within the given precision of the numerical computation. The earlier work with the Feynman-Schwinger representation to scalar-scalar bound states was presented in Ref.\cite{Nieuwenhuis-Tjon}.  
The works of C. Savkli et al. \cite{Savkli:2004bi,Savkli:1999ui,Savkli:2002fj} investigated not only the stability of $\chi^2\phi$-theories but also the cancellation among the vertex corrections, overlapping self-energy, vacuum polarizations, etc. In order to get the full result for two-body bound-states up to the second order in the coupling constant $\alpha$, it has been discussed that it may already be a good approximation just to include the ladder, cross-ladder, stretched-box, and the relevant self-energy corrections and counter-terms. This encouraged us to look into the scalar field model theories with the analytical tools available to us.

An analytically tractable tool for the bound-state problem and also known as the most orthodox tool for dealing with the relativistic two-body problem in quantum field theory is the Bethe-Salpeter formalism \cite{Salpeter:1951sz} utilizing the Green's functions of covariant perturbation theory.  In the ladder approximation of the Bethe-Salpeter formalism the bound-state \cite{Cutkosky:1954ru,Wick:1954eu} problem has been analyzed for the system of two particles interacting with a third scalar particle. However, this approach has fundamental difficulties with the relative time dependence and in systematically including higher-order irreducible kernels such as crossed diagrams and vacuum fluctuations \cite{Brezin:1970zr}.

An alternative approach which can remove these difficulties and restore a systematic perturbative calculation for obtaining higher accuracy is the reformulation of the covariant Bethe-Salpeter equation at equal light-front time, $\tau$ = t + z/c \cite{GPLepage1981,SJBrodsky1982,SJBrodsky1991}. This is equivalent to expressing the Bethe-Salpeter equation in the infinite momentum frame \cite{Dirac:1949cp,Weinberg:1966jm,Susskind:1968rr,Kogut:1969xa,Bjorken:1970ah,Brodsky:1973kb}. The light-front quantization method \cite{Brodsky:1991ir,CRJi:1988n} provides a relativistic Hamiltonian formalism and a Fock-state representation at equal light-front time $\tau$, which retains a lot of the simplicity and utility of the Schr\"{o}dinger nonrelativistic many-body theory \cite{Brodsky:1991ir}. This method not only suppresses the vacuum fluctuations but also systematically includes cross diagrams when higher Fock-state contributions are taken into account. The relativistic Hamiltonian dynamics with equal light-front time $\tau$ has been known as the light-front dynamics (LFD).  

Relativistic two-body bound-states have been analyzed with the light-front formalism of the Bethe-Salpeter approach in the Wick-Cutkosky model \cite{Ji:1985kg,Ji:1985kh}. The light-front ladder approximation in the Wick-Cutkosky model has been extended to the lowest order light-front Tamm-Dancoff approximation, which includes the self-energy corrections and counter-terms \cite{Ji:1994zx}. The cross-ladder and stretched-box up to second order in the coupling constant $\alpha$ have also been included by V.A. Karmanov et al. \cite{Karmanov:2005yg,Carbonell:2006zz}. However, the higher Fock-state contributions due to the particle-antiparticle creation/annihilation process have not yet been included in the cross-ladder contributions \cite{ghkim:dissertation,Sales:2009bp}. Also, the light-front bound-state analyses in the scalar field model theories have largely been limited to the Wick-Cutkosky model which describes the bound-state system with the equal constituent mass and the zero mass of the exchange particle. Therefore, we investigate the contributions of the higher Fock-states due to the particle-antiparticle creation/annihilation process and analyze the two-body bound-state problem in various combinations of masses for the constituent and exchange particles: e.g. $m_1\not=m_2$ and $\lambda\not=0$, when the two constituent particles $\phi_1$ and $\phi_2$ have the masses $m_1$ and $m_2$, respectively, and the exchange particle has the mass $\lambda$. 

In Section 2, we show the light-front formalism of the Bethe-Salpeter equation with the variational method and provide our variational wavefunction suitable for the case of non-zero exchange particle mass $\lambda$. In Section 3, we present the numerical results of the spectrum calculation and relate them to the wavefunction renormalization and the higher Fock-state contribution. The conclusion follows in Section 4.

\section{Formalism}
\subsection{The Bound-State equation in LFD} 
For simplicity, we consider the scalar field model theories which describe bound states of two scalar particles $\phi_1, \phi_2$ with masses $m_1$ and $m_2$ exchanging another scalar particle $\chi$ with mass $\lambda$. While $m_1\not=m_2$ and $\lambda\not=0$ in general, the model with $m_1=m_2$ and $\lambda=0$ has been known as the Wick-Cutkosky model \cite{Wick:1954eu,Cutkosky:1954ru}. The relevant interaction lagrangian is given by 

\begin{equation}
{\cal L}=g(\phi_1\bar{\phi_1}+\phi_2\bar{\phi_2})\chi\,,
\end{equation} 
where g is the coupling constant with the dimension of mass and $\bar{\phi_i}$ ($i=1,2$) is the conjugate field of $\phi_i$. The Bethe-Salpeter equation of this type of model has been reformulated
in LFD and the light-front ladder approximation has been extended to the lowest order light-front Tamm-Dancoff approximation including the self-energy effect \cite{Ji:1994zx}. The cross-ladder and stretched-box up to second order in the coupling constant $\alpha$ have also been included \cite{Karmanov:2005yg,Carbonell:2006zz}. 

\begin{figure}[!h]
\begin{center}
\includegraphics[width=1.0\textwidth]{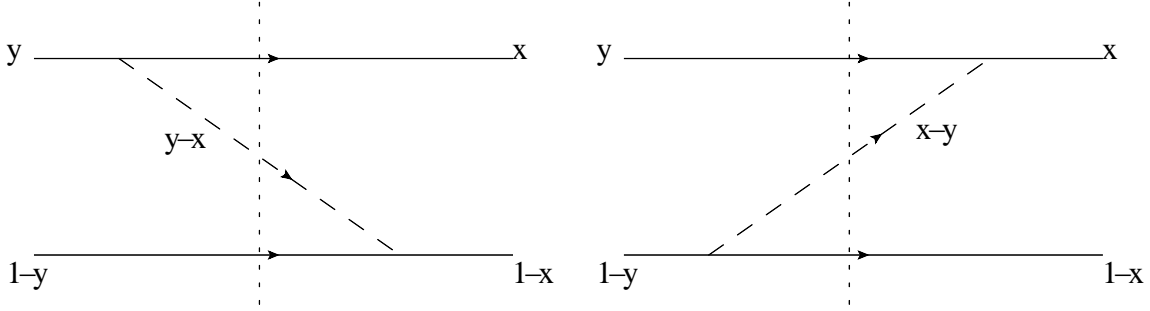}
\caption{Ladder LFD graphs.}
\label{fig:ladderfig}
\end{center}
\end{figure}

\begin{figure}[!h]
\begin{center}
\includegraphics[width=1.0\textwidth]{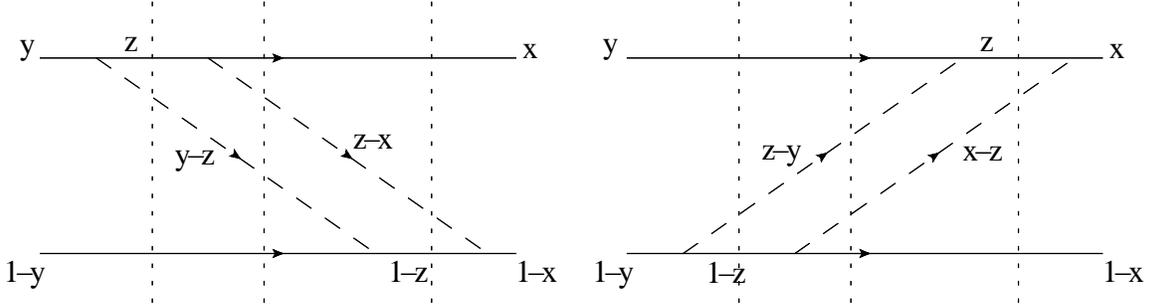}
\caption{Stretched Box LFD graphs.}
\label{stretchfig}
\end{center}
\end{figure}

\begin{figure}[!h]
\begin{center}
\includegraphics[width=1.0\textwidth]{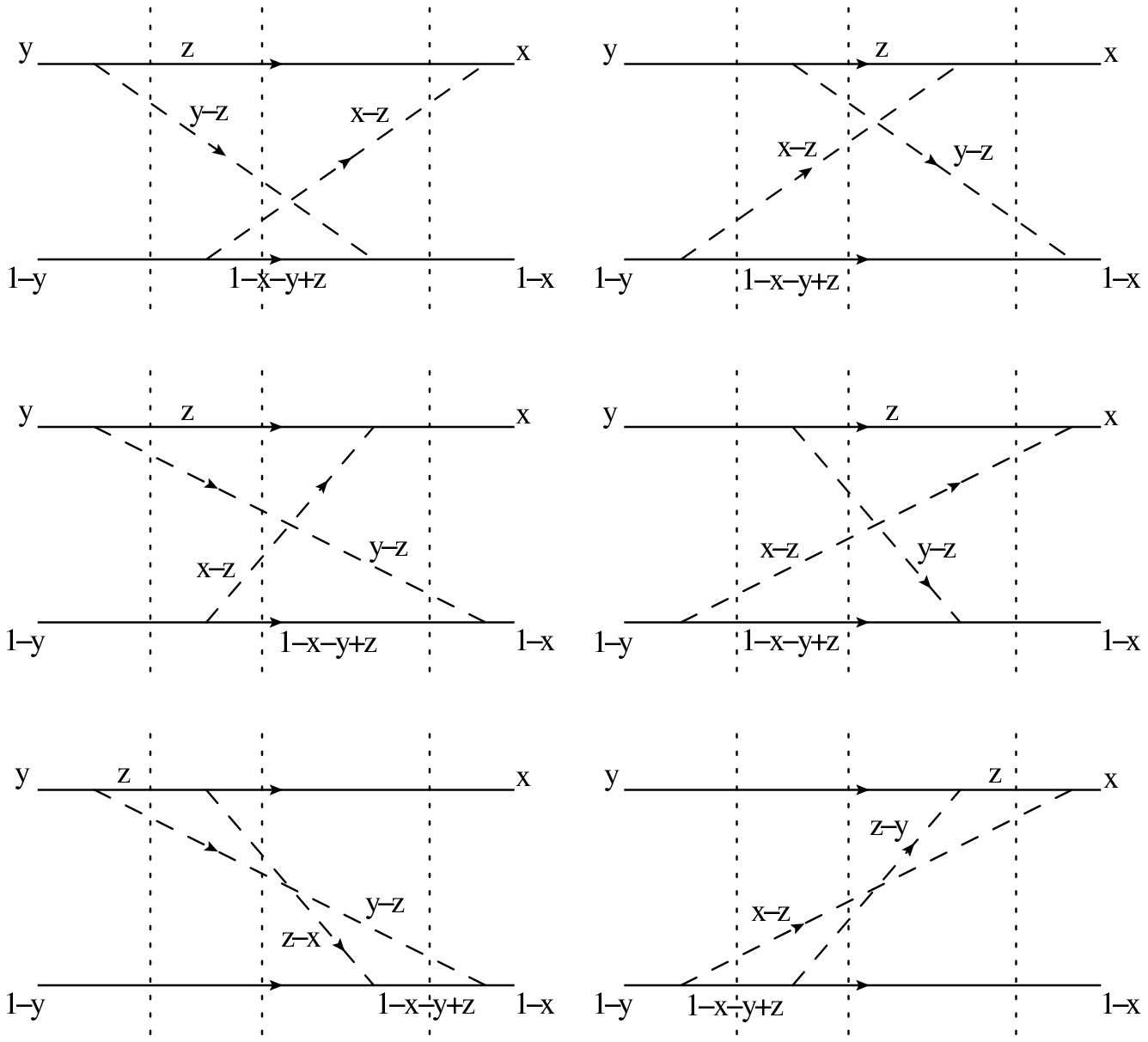}
\caption{Cross Ladder  LFD graphs.}
\label{crossfig}
\end{center}
\end{figure}

\begin{figure}[!h]
\begin{center}
\includegraphics[width=1.0\textwidth]{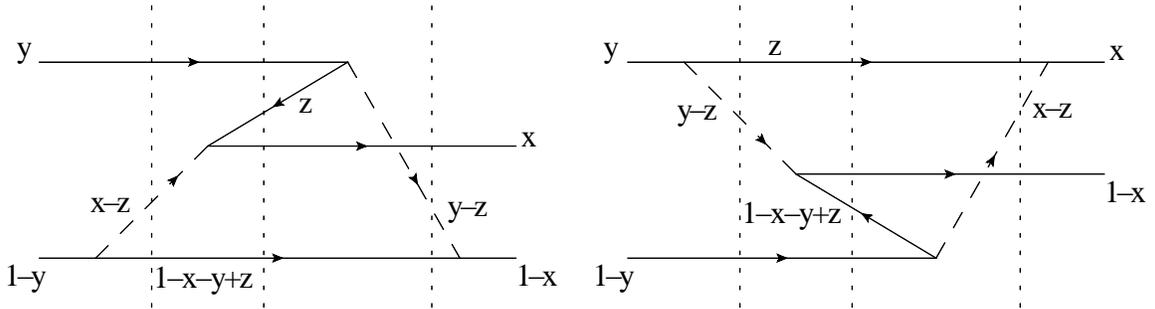}
\caption{Higher Fock LFD graphs.}
\label{hffig}
\end{center}
\end{figure}

In this work, we also go beyond the light-front ladder approximation to include the cross-ladder, stretched-box and the higher Fock-state contributions. In particular, the higher Fock-state contributions (the particle-antiparticle creation/annihilation effect), which has not been numerically analyzed in the previous literature \cite{Carbonell:2006zz,ghkim:dissertation,Sales:2009bp}, are incorporated. The light-front bound-state equation for this theory is given by

\begin{align}
&\left\{M^2 - \frac{\vec{k}_{\perp}^2 + m_1^2}{x}- \frac{\vec{k}_{\perp}^2 + m_2^2}{1-x}-\frac{g^2}{16\pi^2}f (x,\vec{k}_{\perp})\right\} \psi (x,\vec{k}_{\perp})
\nonumber\\
&= \int \frac{dy}{y(1-y)}\frac{d^2\vec{l}_{\perp}}{16\pi^3}K(x,\vec{k}_{\perp};y,\vec{l}_{\perp})\psi(y,\vec{l}_{\perp}) \,, \label{LFBSeq}
\end{align}
where $K(x,\vec{k}_{\perp};y,\vec{l}_{\perp})$ is the kernel of the bound-state equation, $f(x,\vec{k}_{\perp})$ is the self-energy correction, $\psi (x,\vec{k}_{\perp})$ is the light-front wavefunction of the bound state, and $M$ is the mass of the bound state. We denote the light-front longitudinal momentum fraction as $x$, $y$, etc. and the transverse momentum as $\vec{k}_{\perp}$, $\vec{l}_{\perp}$, etc. The kernel $K(x,\vec{k}_{\perp};y,\vec{l}_{\perp})$ is provided up to order $g^4$ including the ladder (L), stretched-box (SB), cross-ladder (CL), and higher-Fock (HF) kernels: i.e.
\begin{align}
K&(x,\vec{k}_{\perp};y,\vec{l}_{\perp})=g^2\, V^{L}(x,\vec{k}_{\perp};y,\vec{l}_{\perp})
\nonumber\\
&+g^4 \, \int dz\frac{d^2\vec{j}_{\perp}}{16\pi^3} \left[V^{SB}(x,\vec{k}_{\perp};y,\vec{l}_{\perp};z,\vec{j}_{\perp})+V^{CL}(x,\vec{k}_{\perp};y,\vec{l}_{\perp};z,\vec{j}_{\perp})+V^{HF}(x,\vec{k}_{\perp};y,\vec{l}_{\perp};z,\vec{j}_{\perp})\right]\,,\label{vkernel}
\end{align}
while the self-energy corrections including counter terms \cite{Ji:1994zx} are given by
\begin{align}
f&(x,\vec{k}_{\perp})=\frac{1}{x}\int_{0}^{1}dz\log\left[1+\frac{x(\frac{\vec{k}_{\perp}^2 + m_1^2}{x}+\frac{\vec{k}_{\perp}^2 + m_2^2}{1-x}-M^2)z(1-z)}{\lambda^2z+m_1^2(1-z)^2}\right]
\nonumber\\
&+\binom{x \leftrightarrow(1-x)}{m_1\leftrightarrow m_2}\,.\label{self-energy}
\end{align}
As the works of C. Savkli et al. \cite{Savkli:2004bi,Savkli:1999ui,Savkli:2002fj} indicated the cancellation among the vertex corrections, overlapping self-energy, vacuum polarizations, etc., we do not include them in the present work but limit here only up to the four-body Fock states in LFD. We will be content with just taking into account the kernel $K(x,\vec{k}_{\perp};y,\vec{l}_{\perp})$ in Eq.~(\ref{vkernel}) and the self-energy correction term $f(x,\vec{k}_{\perp})$ in Eq.~(\ref{self-energy}). While the ladder kernel ($V^{L}$) is given by
\begin{align}
V^{L}&(x,\vec{k}_{\perp},y,\vec{l}_{\perp})=\frac{\theta(x-y)}{(x-y)}\frac{1}{M^2 - \frac{\vec{l}_{\perp}^2 + m_1^2}{y} - \frac{(\vec{k}_{\perp}-\vec{l}_{\perp})^2 + \lambda^2}{x-y}- \frac{\vec{k}_{\perp}^2 + m_2^2}{1-x} }
\nonumber\\
&+\binom{x \leftrightarrow y}{\vec{k}_{\perp}  \leftrightarrow \vec{l}_{\perp}}\,,
\end{align}
the stretched-box ($V^{SB}$), cross-ladder ($V^{CL}$), and higher-Fock ($V^{HF}$) kernels can be written in the form:
\begin{align}
&V^{SB}(x,\vec{k}_{\perp};y,\vec{l}_{\perp};z,\vec{j}_{\perp})=\frac{1}{z(x-z)(z-y)(1-z)}\sum_{i=1,2}V_i\,,\\
&V^{CL}(x,\vec{k}_{\perp};y,\vec{l}_{\perp};z,\vec{j}_{\perp})=\frac{1}{z(x-z)(z-y)(1-x-y+z)}\sum_{i=3}^{8}V_i\,,\\
&V^{HF}(x,\vec{k}_{\perp};y,\vec{l}_{\perp};z,\vec{j}_{\perp})=\frac{1}{z(x-z)(z-y)(1-x-y+z)}\left[V_A+V_B\right]\,,
\end{align}
where $V_1,V_2,V_3,V_4,V_5,V_6,V_7,V_8,V_A$, and $V_B$ are presented in Appendix A. The corresponding diagrams are shown in Figs.~\ref{fig:ladderfig} -- \ref{hffig}. In particular, the particle-antiparticle creation/annihilation process which was not included in the previous cross-ladder analysis \cite{Carbonell:2006zz} is shown in  Fig.~\ref{hffig}. We denote this contribution as $V^{HF}$.

\newpage
\subsection{Spectrum Calculation with Variational Method} 
In order to solve Eq.~(\ref{LFBSeq}) numerically, we utilize a variational principle for a dimensionless coupling constant $\alpha$ given by $\alpha=g^2/(16\pi m_1m_2)$. Taking the expectation values with a variational wavefunction in Eq.~(\ref{LFBSeq}), we get the quadratic equation in terms of $\alpha$:
\begin{equation}
\left\langle M^2 - \frac{\vec{k}_{\perp}^2 + m_1^2}{x}- \frac{\vec{k}_{\perp}^2 + m_2^2}{1-x}\right\rangle 
=\alpha \left\langle V^L\right\rangle+\alpha\left\langle\frac{m_1m_2}{\pi}f\right\rangle+\alpha^2 \left\langle V^{SB}+V^{CL}+V^{HF}\right\rangle\,,
\end{equation}
where more explicitly each of the expectation values are given by
\begin{align}
\left\langle M^2 - \frac{\vec{k}_{\perp}^2 + m_1^2}{x}- \frac{\vec{k}_{\perp}^2 + m_2^2}{1-x}\right\rangle=\int \frac{dx}{x(1-x)}\frac{d^2\vec{k}_{\perp}}{16\pi^3}\psi^{\dagger} (x,\vec{k}_{\perp}) \left\{M^2 - \frac{\vec{k}_{\perp}^2 + m_1^2}{x}- \frac{\vec{k}_{\perp}^2 + m_2^2}{1-x}\right\} \psi(x,\vec{k}_{\perp})\,,
\end{align}
\begin{align}
\left\langle\frac{m_1m_2}{\pi}f\right\rangle=\int \frac{dx}{x(1-x)}\frac{d^2\vec{k}_{\perp}}{16\pi^3}\psi^{\dagger} (x,\vec{k}_{\perp})\frac{m_1m_2}{\pi}f (x,\vec{k}_{\perp})\psi(x,\vec{k}_{\perp})\,,\label{expvalself}
\end{align}
\begin{align}
\left\langle V^L\right\rangle=16\pi m_1m_2\int \frac{dx}{x(1-x)}\frac{d^2\vec{k}_{\perp}}{16\pi^3}\frac{dy}{y(1-y)}\frac{d^2\vec{l}_{\perp}}{16\pi^3}\psi^{\dagger} (x,\vec{k}_{\perp})V^{L}(x,\vec{k}_{\perp};y,\vec{l}_{\perp})\psi(y,\vec{l}_{\perp})\,,
\end{align}
\begin{align}
\left\langle V^{SB}\right\rangle=(16\pi m_1m_2)^2\int \frac{dx}{x(1-x)}\frac{d^2\vec{k}_{\perp}}{16\pi^3}&\int \frac{dy}{y(1-y)}\frac{d^2\vec{l}_{\perp}}{16\pi^3}\int dz\frac{d^2\vec{j}_{\perp}}{16\pi^3} \psi^{\dagger} (x,\vec{k}_{\perp})
\nonumber\\
&\times V^{SB}(x,\vec{k}_{\perp};y,\vec{l}_{\perp};z,\vec{j}_{\perp})\psi(y,\vec{l}_{\perp})\,,
\end{align}
\begin{align}
\left\langle V^{CL}\right\rangle= (16\pi m_1m_2)^2\int \frac{dx}{x(1-x)}\frac{d^2\vec{k}_{\perp}}{16\pi^3}&\int \frac{dy}{y(1-y)}\frac{d^2\vec{l}_{\perp}}{16\pi^3}\int dz\frac{d^2\vec{j}_{\perp}}{16\pi^3} \psi^{\dagger} (x,\vec{k}_{\perp})
\nonumber\\
&\times V^{CL}(x,\vec{k}_{\perp};y,\vec{l}_{\perp};z,\vec{j}_{\perp})\psi(y,\vec{l}_{\perp})\,,
\end{align}
\begin{align}
\left\langle V^{HF}\right\rangle=(16\pi m_1m_2)^2\int \frac{dx}{x(1-x)}\frac{d^2\vec{k}_{\perp}}{16\pi^3}&\int \frac{dy}{y(1-y)}\frac{d^2\vec{l}_{\perp}}{16\pi^3}\int dz\frac{d^2\vec{j}_{\perp}}{16\pi^3} \psi^{\dagger} (x,\vec{k}_{\perp})
\nonumber\\
&\times V^{HF}(x,\vec{k}_{\perp};y,\vec{l}_{\perp};z,\vec{j}_{\perp})\psi(y,\vec{l}_{\perp})\,.\label{expvalhf}
\end{align}
For fixed binding energy $B=m_1+m_2-M$, we solve the quadratic equation of the coupling constant $\alpha$ and minimize the corresponding expectation values to determine the optimum relation between the binding energy $B$ and the coupling constant $\alpha$:
\begin{equation}
\alpha=\frac{-\left\langle V^L+\frac{m_1m_2}{\pi}f \right\rangle+\sqrt{\left\langle V^L+\frac{m_1m_2}{\pi}f \right\rangle^2+4\left\langle M^2 - \frac{\vec{k}_{\perp}^2 + m_1^2}{x}- \frac{\vec{k}_{\perp}^2 + m_2^2}{1-x}\right\rangle \left\langle V^{CL}+V^{SB}+V^{HF} \right\rangle}}{ 2\left\langle V^{CL}+V^{SB}+V^{HF} \right\rangle}\,.\label{alphaeq}
\end{equation}
The minimum of $\alpha$ is found by varying the parameters in the trial wavefunction $\psi(x,\vec{k}_{\perp})$. A judicious choice of trial wavefunction is important to get the closest result to the true minimum of $\alpha$. To achieve this goal, we consider available exact solutions in some limiting cases, e.g. the solution of the non-relativistic Schr\"{o}dinger equation for the Coulomb interaction. We also take into account the relation of the light-front bound-state equation to the covariant Bethe-Salpeter equation, since some analytic spectral functions for the solution of the Bethe-Salpeter equation are known in both weak and strong binding limits \cite{Nakanishi:1969ph}. For the $1S$ state, we take the variational wavefunction parameterized by 
\begin{align}
\psi(x,\vec{k}_{\perp}) = &\frac{N_{1s}}{\left[C^2 - \frac{\vec{k}_{\perp}^2 + m_1^2}{x}- \frac{\vec{k}_{\perp}^2 + m_2^2}{1-x}\right]}\times  \nonumber \\
&\frac{1}{\left[C^2-\frac{\vec{k}_{\perp}^2 + m_1^2}{x}- \frac{\vec{k}_{\perp}^2 + m_2^2}{1-x}-4\lambda\left(\lambda +\sqrt{2(m_1^2+m_2^2)-\frac{(m_1^2-m_2^2)^2}{(\frac{\vec{k}_{\perp}^2+m_1^2}{x}+\frac{\vec{k}_{\perp}^2+m_2^2}{1-x})}-C^2}\right)\right]{(1+|2x-1|)}}\,,\label{psi}
\end{align}
where the normalization constant $N_{1s}$ cancels in the ratio given by Eq.~(\ref{alphaeq}) and $C$ is the variational parameter. The factor $(1+|2x-1|)$ in the denominator of Eq.~(\ref{psi}) stems from the weak binding spectral function of the Bethe-Salpeter solution as shown in \cite{Brodsky:1984vp}. Eq.~(\ref{psi}) is the variational wavefunction suitable for the case of non-zero exchange particle mass $\lambda$. We tried the variational wavefunction without the $\lambda$ term in Eq.~(\ref{psi}) as well as other trial wavefunctions different from Eq.~(\ref{psi}), but found that the obtained expectation values of $\alpha$ were all larger than what we get using Eq.~(\ref{psi}). This convinced us that the wavefunction given by Eq.~(\ref{psi}) is an improvement over any other variational wavefunctions that we have considered. The origin of  Eq.~(\ref{psi}) may be attributed to the exact solution of the Schr\"{o}dinger equation with the Hulthen potential \cite{hulthen:encyclopedia,hulthen:1941,hulthen:1942}. Although the Schr\"{o}dinger equation with the Yukawa potential has not yet been analytically solved, the Hulthen potential behaves like the Yukawa potential for small values of $r$ and the exact solutions have been obtained for the Schr\"{o}dinger equation with the Hulthen potential. With the form of the variational wavefunction given by Eq.~(\ref{psi}), we are well equipped to compute mass spectra even for the case of $\lambda\not=0$.

\subsection{Analytic Calculation of Spectrum and Non-Relativistic Limit} 
A simple analytic relation between the coupling constant $\alpha$ and the binding energy $B$ may be attained by following the method presented in Refs.~\cite{Ji:1994zx,Ji:1985kg}. For example, if we take $C=M$ in Eq.~(\ref{psi}) for the case of $m_1=m_2=m$ and $\lambda=0$, we get
\begin{eqnarray}
\psi(x,\vec{k}_{\perp})= \frac{N}{\left[M^2 - \frac{\vec{k}_{\perp}^2 + m^2}{x(1-x)}\right]^2 {(1+|2x-1|)}}\,.\label{psiequalmasslambdazero}
\end{eqnarray}
Without the factor $(1+|2x-1|)$ in the denominator of Eq.~(\ref{psiequalmasslambdazero}), this wavefunction corresponds to the $1S$ state wavefunction of the Schr\"{o}dinger equation with the Coulomb potential. Eq.~(\ref{psiequalmasslambdazero}) was used to find the analytic expression for the binding energy of the $1S$ state of the ladder approximation originally in \cite{Ji:1985kg,Ji:1985kh}. Including the self-energy correction, the following formula was obtained \cite{Ji:1994zx}:
\begin{equation}
\frac{\pi}{\alpha}=\{(u-2)/(u-1)^{1/2}\}\{\frac{1}{2}\pi-\tan^{-1}((u-1)^{-1/2})\}+\log(4/u)+\frac{2u}{u-2}\log\frac{2}{u}\,,\label{pioveralphaeqlmass}
\end{equation}
where $\alpha=g^2/(16\pi m^2)$ and $u=m^2/(m^2-M^2/4)$. In the zero-binding limit, $u\to\infty$, Eq.~(\ref{pioveralphaeqlmass}) leads to $\frac{\pi}{\alpha}=\frac{\pi}{2}\sqrt{u}$, which is the well-known Balmer formula for the $1S$ state: i.e.  
\begin{equation}
2m-M=m\alpha^2/4\,.\label{balmerformulaequalmass}
\end{equation}
Similarly, following the method presented in \cite{Ji:1994zx} as well as in \cite{Ji:1985kg}, we obtain the following relation for $m_1=m_2=m$ and $\lambda\not=0$:
\begin{align}
\frac{\pi}{\alpha}=8u^2\int_{4}^{\infty}dt\frac{(t-2)\left(1+\frac{\lambda}{m}\sqrt{u}\right)^2\log\left\{\frac{u\left[4+t(t-4)\frac{\lambda^2}{m^2}\right]}{4(t-4)+u\left[4+(t-2)(t-4)\frac{\lambda^2}{m^2}\right]}\right\}}{\left(\frac{\lambda^2}{m^2}u-2\right)\left[t(t-4)+4u\right]\left[t(t-4)\left(1+\frac{\lambda}{m}\sqrt{u}\right)^2+4u\right]}-u\int_{0}^{1}dz\log\left[1+\frac{2z(1-z)}{\frac{\lambda^2}{(m^2-\frac{M^2}{4})}z+u(1-z)^2}\right]\,.\label{pioveralphaeqlmassnonzerolambda}
\end{align}
For $\lambda=0$, Eq.~(\ref{pioveralphaeqlmassnonzerolambda}) can be integrated and reduced to Eq.~(\ref{pioveralphaeqlmass}). We will discuss these relations further in the zero-binding energy region when we present our numerical results in Section 3.

\subsection{Wavefunction Renormalization and Probabilities of Lowest and Higher Fock-States} 
The light-front quantization method provides a description of a bound-state $|B\rangle$ in a Fock-state representation at equal $\tau$: 
\begin{equation}
|B\rangle=\langle\phi\bar{\phi}|B\rangle|\phi\bar{\phi}\rangle+\langle\phi\bar{\phi}\chi|B\rangle|\phi\bar{\phi}\chi\rangle+\langle\phi\bar{\phi}\chi\chi|B\rangle|\phi\bar{\phi}\chi\chi\rangle+ \cdots\,,\label{boundstateb}
\end{equation}
where the light-front two-body wavefunction $\psi(x,\vec{k}_{\perp})$ corresponds to the lowest Fock-state amplitude $\langle \phi\bar{\phi}|B\rangle$. The probability to find the two-body state $|\phi\bar{\phi}\rangle$ is given by the integration of the wavefunction squared: 
\begin{equation}
P_{\rm Low}=\int \frac{dx}{x(1-x)}\frac{d^2\vec{k}_{\perp}}{16\pi^3}|\psi(x,\vec{k}_{\perp})|^2\,,
\end{equation}
where the subscript ``$\rm Low$" of $P_{\rm Low}$ is introduced because the two-body state is the lowest Fock-state. The probability to find the higher Fock-states can be obtained by differentiating the kernel $K(x,\vec{k}_{\perp};y,\vec{l}_{\perp})$ in Eq.~(\ref{vkernel}) with respect to $M^2$ as discussed in Refs.~\cite{Ji:1994zx,Lepage:1980fj}: i.e.
\begin{align}
P_{\rm High}&=\left\langle -\frac{\partial K}{\partial M^2}\right\rangle
\nonumber\\
&=-\,\int \frac{dx}{x(1-x)}\frac{d^2\vec{k}_{\perp}}{16\pi^3}\frac{dy}{y(1-y)}\frac{d^2\vec{l}_{\perp}}{16\pi^3}\psi^{\dagger} (x,\vec{k}_{\perp})\left(\frac{\partial K}{\partial M^2}\right)\psi(y,\vec{l}_{\perp})\,,
\end{align}
where the subscript ``$\rm High$" of $P_{\rm High}$ indicates the contributions from the higher Fock-states such as three and four bodies through $V^{L}$, $V^{SB}$, $V^{CL}$, and $V^{HF}$.  Including the self-energy correction, we may also renormalize the two-body wavefunction as discussed in Ref.~\cite{Brodsky:1984vp}: i.e.
\begin{equation}
\tilde{P}_{\rm Low}=\int \frac{dx}{x(1-x)}\frac{d^2\vec{k}_{\perp}}{16\pi^3}\left[1-(\frac{\alpha m_1m_2}{\pi})\frac{\partial f (x,\vec{k}_{\perp})}{\partial M^2}\right]|\tilde{\psi}(x,\vec{k}_{\perp})|^2\,,\label{plowwithself}
\end{equation}
where $\tilde{\psi}(x,\vec{k}_{\perp})$ is the wavefunction including the self-energy correction. The probability to find the higher Fock-states including the self-energy correction is given by
\begin{align}
\tilde{P}_{\rm High}&=\left\langle -\frac{\partial K}{\partial M^2}\right\rangle
\nonumber\\
&=-\,\int \frac{dx}{x(1-x)}\frac{d^2\vec{k}_{\perp}}{16\pi^3}\frac{dy}{y(1-y)}\frac{d^2\vec{l}_{\perp}}{16\pi^3}\tilde{\psi}^{\dagger} (x,\vec{k}_{\perp})\left(\frac{\partial K}{\partial M^2}\right)\tilde{\psi}(y,\vec{l}_{\perp})\,.\label{phighwithself}
\end{align}
Since the kernel $K$ can be decomposed into the contributions from $V^{L}$, $V^{SB}$, $V^{CL}$, and $V^{HF}$, the probabilities $P_{\rm Low}$, $\tilde{P}_{\rm Low}$, $P_{\rm High}$, and $\tilde{P}_{\rm High}$ can also be decomposed into the corresponding contributions. For example, up to the ladder and cross-ladder, we may define the following probabilities\footnote{see Appendix B for the full expression of $P_{\rm Low}$, $\tilde{P}_{\rm Low}$, $P_{\rm High}$, and $\tilde{P}_{\rm High}$.}:
\begin{align}
P_{\rm Low}^L=&\int \frac{dx}{x(1-x)}\frac{d^2\vec{k}_{\perp}}{16\pi^3}|\psi_L(x,\vec{k}_{\perp})|^2\,,
\nonumber\\
P_{\rm Low}^{L+CL}=&\int \frac{dx}{x(1-x)}\frac{d^2\vec{k}_{\perp}}{16\pi^3}|\psi_{L+CL}(x,\vec{k}_{\perp})|^2\,,
\nonumber\\
\tilde{P}_{\rm Low}^{L}=&\int \frac{dx}{x(1-x)}\frac{d^2\vec{k}_{\perp}}{16\pi^3}\left[1-(\frac{\alpha m_1m_2}{\pi})\frac{\partial f (x,\vec{k}_{\perp})}{\partial M^2}\right]|\tilde{\psi}_{L}(x,\vec{k}_{\perp})|^2\,,
\nonumber\\
\tilde{P}_{\rm Low}^{L+CL}=&\int \frac{dx}{x(1-x)}\frac{d^2\vec{k}_{\perp}}{16\pi^3}\left[1-(\frac{\alpha m_1m_2}{\pi})\frac{\partial f (x,\vec{k}_{\perp})}{\partial M^2}\right]|\tilde{\psi}_{L+CL}(x,\vec{k}_{\perp})|^2\,,
\nonumber\\
P_{\rm High}^{L}=&\left\langle -\frac{\partial(\alpha V^{L})}{\partial M^2}\right\rangle
\nonumber\\
=&-g^2\,\int \frac{dx}{x(1-x)}\frac{d^2\vec{k}_{\perp}}{16\pi^3}\int \frac{dy}{y(1-y)}\frac{d^2\vec{l}_{\perp}}{16\pi^3}\psi^{\dagger}_L (x,\vec{k}_{\perp})\Big(\frac{\partial V^{L}}{\partial M^2}\Big)\psi_L(y,\vec{l}_{\perp})\,,
\nonumber\\
P_{\rm High}^{L+CL}=&\left\langle -\frac{\partial(\alpha V^{L}+\alpha^2 V^{CL})}{\partial M^2}\right\rangle
\nonumber\\
=&-g^2\,\int \frac{dx}{x(1-x)}\frac{d^2\vec{k}_{\perp}}{16\pi^3}\int \frac{dy}{y(1-y)}\frac{d^2\vec{l}_{\perp}}{16\pi^3}\psi^{\dagger}_{L+CL} (x,\vec{k}_{\perp})\left(\frac{\partial V^{L}}{\partial M^2}\right)\psi_{L+CL}(y,\vec{l}_{\perp})
\nonumber\\
&-g^4\,\int \frac{dx}{x(1-x)}\frac{d^2\vec{k}_{\perp}}{16\pi^3}\int \frac{dy}{y(1-y)}\frac{d^2\vec{l}_{\perp}}{16\pi^3}\int dz\frac{d^2\vec{j}_{\perp}}{16\pi^3}\psi^{\dagger}_{L+CL} (x,\vec{k}_{\perp})\left(\frac{\partial V^{CL}}{\partial M^2}\right)\psi_{L+CL}(y,\vec{l}_{\perp})\,,
\nonumber\\
\tilde{P}_{\rm High}^{L}=& \left\langle -\frac{\partial(\alpha V^{L})}{\partial M^2}\right\rangle
\nonumber\\
&=-g^2\,\int \frac{dx}{x(1-x)}\frac{d^2\vec{k}_{\perp}}{16\pi^3}\int \frac{dy}{y(1-y)}\frac{d^2\vec{l}_{\perp}}{16\pi^3}\tilde{\psi}^{\dagger}_{L} (x,\vec{k}_{\perp})\left(\frac{\partial V^{L}}{\partial M^2}\right)\tilde{\psi}_{L}(y,\vec{l}_{\perp})\,,
\nonumber\\
\tilde{P}_{\rm High}^{L+CL}=&\left\langle -\frac{\partial(\alpha V^{L}+\alpha^2 V^{CL})}{\partial M^2}\right\rangle
\nonumber\\
=&-g^2\,\int \frac{dx}{x(1-x)}\frac{d^2\vec{k}_{\perp}}{16\pi^3}\int \frac{dy}{y(1-y)}\frac{d^2\vec{l}_{\perp}}{16\pi^3}\tilde{\psi}^{\dagger}_{L+CL} (x,\vec{k}_{\perp})\Big(\frac{\partial V^{L}}{\partial M^2}\Big)\tilde{\psi}_{L+CL}(y,\vec{l}_{\perp})
\nonumber\\
&-g^4\,\int \frac{dx}{x(1-x)}\frac{d^2\vec{k}_{\perp}}{16\pi^3}\int \frac{dy}{y(1-y)}\frac{d^2\vec{l}_{\perp}}{16\pi^3}\int dz\frac{d^2\vec{j}_{\perp}}{16\pi^3}\tilde{\psi}^{\dagger}_{L+CL} (x,\vec{k}_{\perp})\Big(\frac{\partial V^{CL}}{\partial M^2}\Big)\tilde{\psi}_{L+CL}(y,\vec{l}_{\perp})\,,
\end{align}
where $\psi_{L}(x,\vec{k}_{\perp})$ is the two-body wavefunction obtained by including only the ladder kernel $V^{L}$ while $\psi_{L+CL}(x,\vec{k}_{\perp})$ and $\tilde{\psi}_{L}(x,\vec{k}_{\perp})$ are the wavefunctions including the cross-ladder kernel $V^{CL}$ and the self-energy correction term $f(x,\vec{k}_{\perp})$, respectively, in addition to $V^{L}$. Likewise, we define $P_{\rm Low}^{L+CL+SB}$, $\tilde{P}_{\rm Low}^{L+CL+SB}$, $P_{\rm High}^{L+CL+SB}$, and $\tilde{P}_{\rm High}^{L+CL+SB}$ by adding the contribution from $V^{SB}$. By adding the contribution from $V^{HF}$ to this, we may also define $P_{\rm Low}^{L+CL+SB+HF}$, $\tilde{P}_{\rm Low}^{L+CL+SB+HF}$, $P_{\rm High}^{L+CL+SB+HF}$, and $\tilde{P}_{\rm High}^{L+CL+SB+HF}$. Here, the factor $16\pi m_1m_2$ in the relation between $\alpha$ and $g^2$ has been compensated by the same factor included in the definition of the expectation values given by Eqs.~(\ref{expvalself}) -- (\ref{expvalhf}). We note that the spectrum of the bound-state is intimately correlated with the corresponding bound-state wavefunction. Thus, it is interesting to  compute the probabilities $P_{\rm Low}$, $\tilde{P}_{\rm Low}$, $P_{\rm High}$, and $\tilde{P}_{\rm High}$ with the specific contributions from the kernel and/or the self-energy correction and discuss the correlation of these wavefunction-related observables with the results of the spectrum. We present the probability ratios such as $P_{\rm High}^{L}/P_{\rm Low}^{L}$, $P_{\rm High}^{L+CL}/P_{\rm Low}^{L+CL}$, $\tilde{P}_{\rm High}^{L}/\tilde{P}_{\rm Low}^{L}$, etc. in the following section of Numerical Results.
\section{Numerical Results}
\subsection{Spectrum} 
We present the results for the binding energy $B=m_1+m_2-M$ as a function of the coupling constant $\alpha=g^2/(16\pi m_1m_2)$. Since we consider in general the $m_1\not=m_2$ and $\lambda\not=0$ cases as well as the special cases of $m_1=m_2$ or $\lambda=0$, we present the numerical results of representative examples ($\lambda/m_2=0, 0.15, 0.5, 1.0, 2.5$) with $\beta\equiv m_1/m_2=1$ and $5.446\times 10^{-4}$ in units  where $m_2=1$. Whether the range of the interaction between the two constituents of the bound-state is short or long depends on the value of $\lambda$: e.g. the Yukawa interaction with $\lambda\not=0$ or the Coulomb interaction with $\lambda=0$. Also, the value of $\beta$ is linked to the modeling of the bound-state: e.g. positronium or deutron system with $\beta=1$, and hydrogen atom with $\beta=5.446\times10^{-4}$. 

\begin{figure}
\begin{center}
\includegraphics[width=1.0\linewidth]{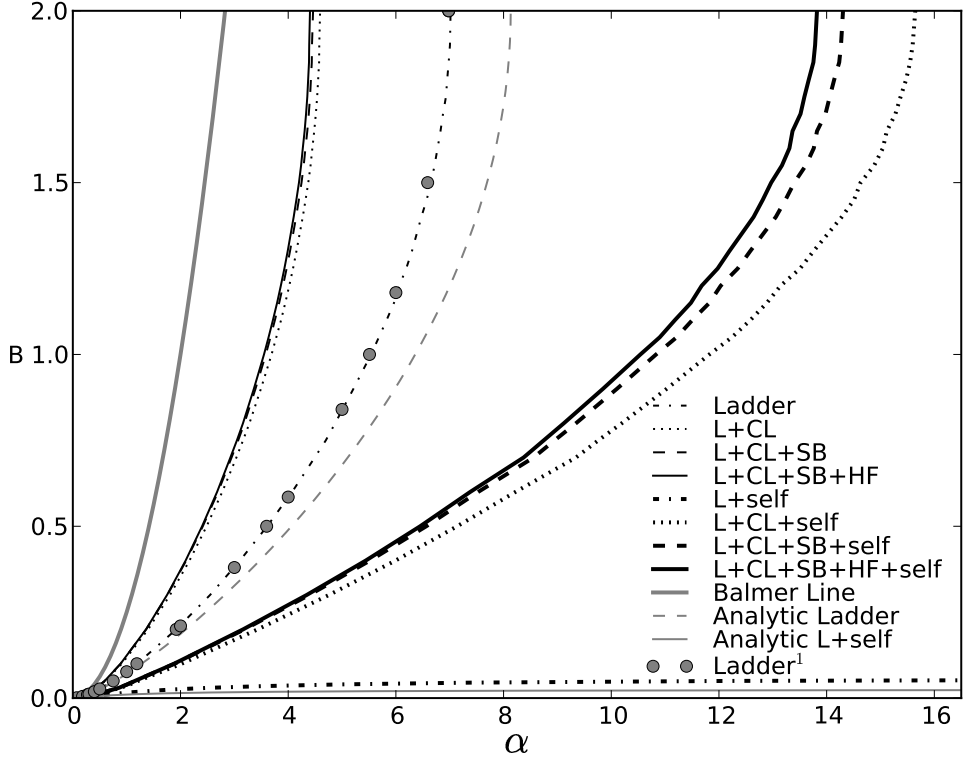}
\caption{$\beta=1$, $\lambda=0$. The numerical results of the ladder, $L+CL$, $L+CL+SB$, and $L+CL+SB+HF$ with and without the self-energy corrections obtained by the variational principle. The ladder and L+self are compared with the corresponding analytic results from Eq.~(\ref{pioveralphaeqlmass}). The Balmer line is shown as the non-relativistic result from Eq.~(\ref{balmerformulaequalmass}). The ladder is also compared with the previous data from M.\ Mangin-Brinet et al. $^1$(M.\ Mangin-Brinet and J.\ Carbonell \cite{ManginBrinet:1999fh}.)}
\label{beta1lambda0includeeverything}
\end{center}
\end{figure}

Fig.~\ref{beta1lambda0includeeverything} shows the case of $\beta=1, \lambda=0$. The numerical results of the ladder, $L+CL$, $L+CL+SB$, and $L+CL+SB+HF$ with and without the self-energy corrections obtained by the variational principle are presented and compared with the available numerical or analytic results. The non-relativistic result given by Eq.~(\ref{balmerformulaequalmass}) is shown as the Balmer Line (thick grey solid line) in the far left of this figure, while the analytic result (Analytic $L+${\it self}) given by Eq.~(\ref{pioveralphaeqlmass}) is shown as a thin grey solid line at the bottom of the figure. In the middle between these two curves (Balmer Line and Analytic $L+${\it self}), our ladder result (thin black dot-dashed line) is compared with the previous numerical results (solid circles) provided by M.\ Mangin-Brinet and J.\ Carbonell \cite{ManginBrinet:1999fh}. Although our method of computation is different from theirs \cite{ManginBrinet:1999fh}, our ladder results are in good agreement with their ladder results. It lends confidence to our variational calculation. Just next to the ladder result on the right, the analytic result without the self-energy correction, i.e. without the last term in Eq.~(\ref{pioveralphaeqlmass}), is shown as a thin grey dashed line. The analytic result agrees with the numerical result in the weak binding limit. Similarly, the analytic result including the self-energy effect (Analytic $L+${\it self}) shown at the bottom of the figure is comparable to the corresponding numerical results ($L+${\it self}) drawn as a thick black dot-dashed line just above the analytic result.  Between the Balmer line and the ladder result, one can see the three curves of $L+CL$, $L+CL+SB$ and $L+CL+SB+HF$ from right to left. Also, between the ladder result and the L+self result at the bottom, one can see the three corresponding curves including the self-energy effect. These results indicate that the effect of the self-energy term is highly repulsive while the contribution from the cross-ladder shows a large attractive effect. The difference between the $L+${\it self} and $L+CL+${\it self} is more significant than the difference between the ladder and $L+CL$, which indicates that the attractive effect from the cross-ladder kernel is more significant when the self-energy corrections are included. We note that the particle-antiparticle creation/annihilation contribution from $V^{HF}$ as well as the stretched-box contribution from $V^{SB}$ yield attractive effects and become larger as the coupling constant $\alpha$ increases. Although the effect of the particle-antiparticle creation/annihilation is very small in this case ($\beta=1, \lambda=0$), it appears more noticeable when the self-energy corrections are included as one can see in the difference between the thick black solid and thick black dashed lines. 

\begin{figure}
\begin{minipage}{0.5\textwidth}
\begin{center}
\includegraphics[width=1.0\linewidth]{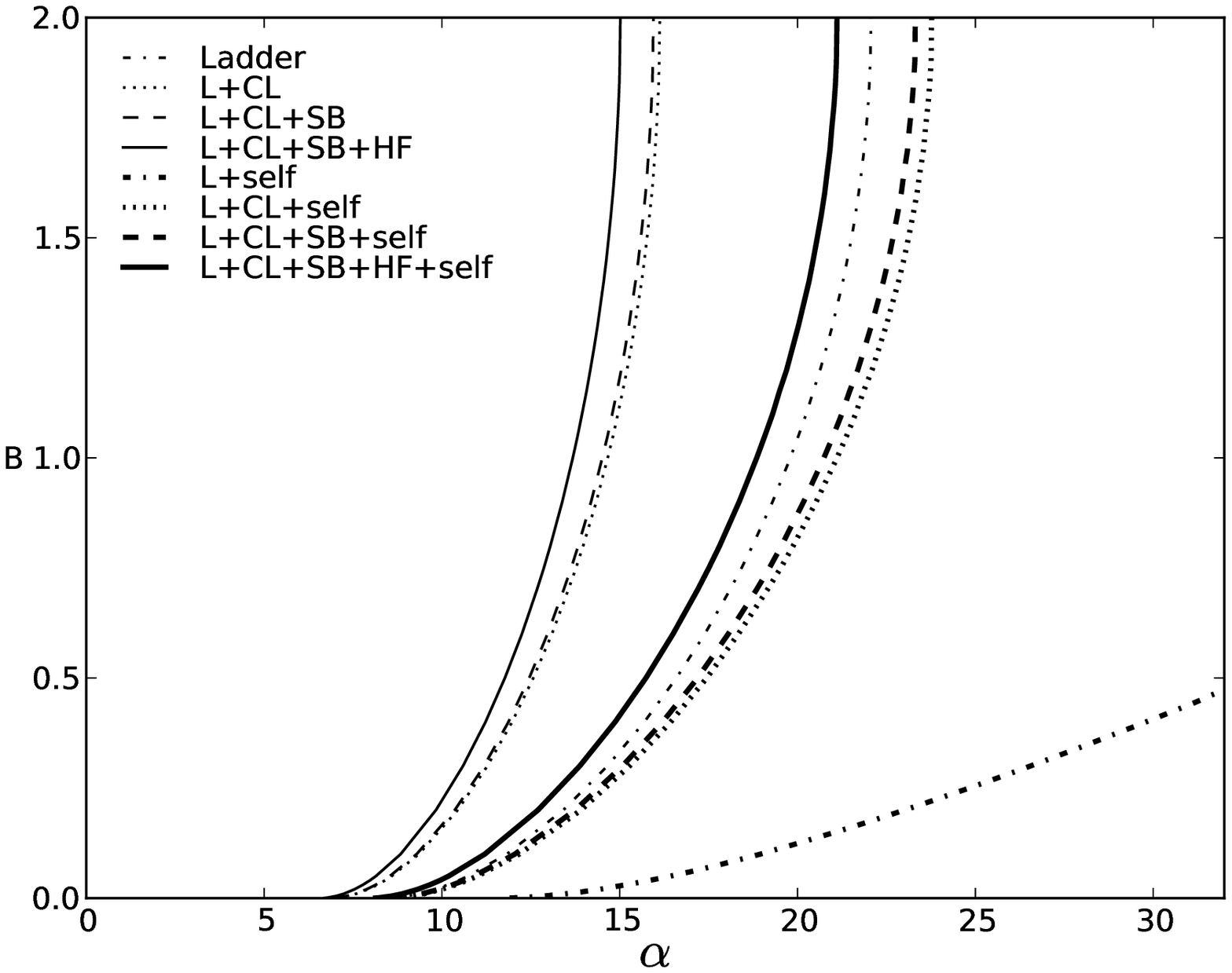}
\caption{$\beta=1$, $\lambda=2.5$. The numerical results of the ladder, $L+CL$, $L+CL+SB$, and $L+CL+SB+HF$ with and without the self-energy corrections obtained by the variational principle.}
\label{beta1lambda25}
\end{center}
\end{minipage}
\hspace{0.0\linewidth}
\begin{minipage}{0.5\textwidth}
\begin{center}
\includegraphics[width=1.0\linewidth]{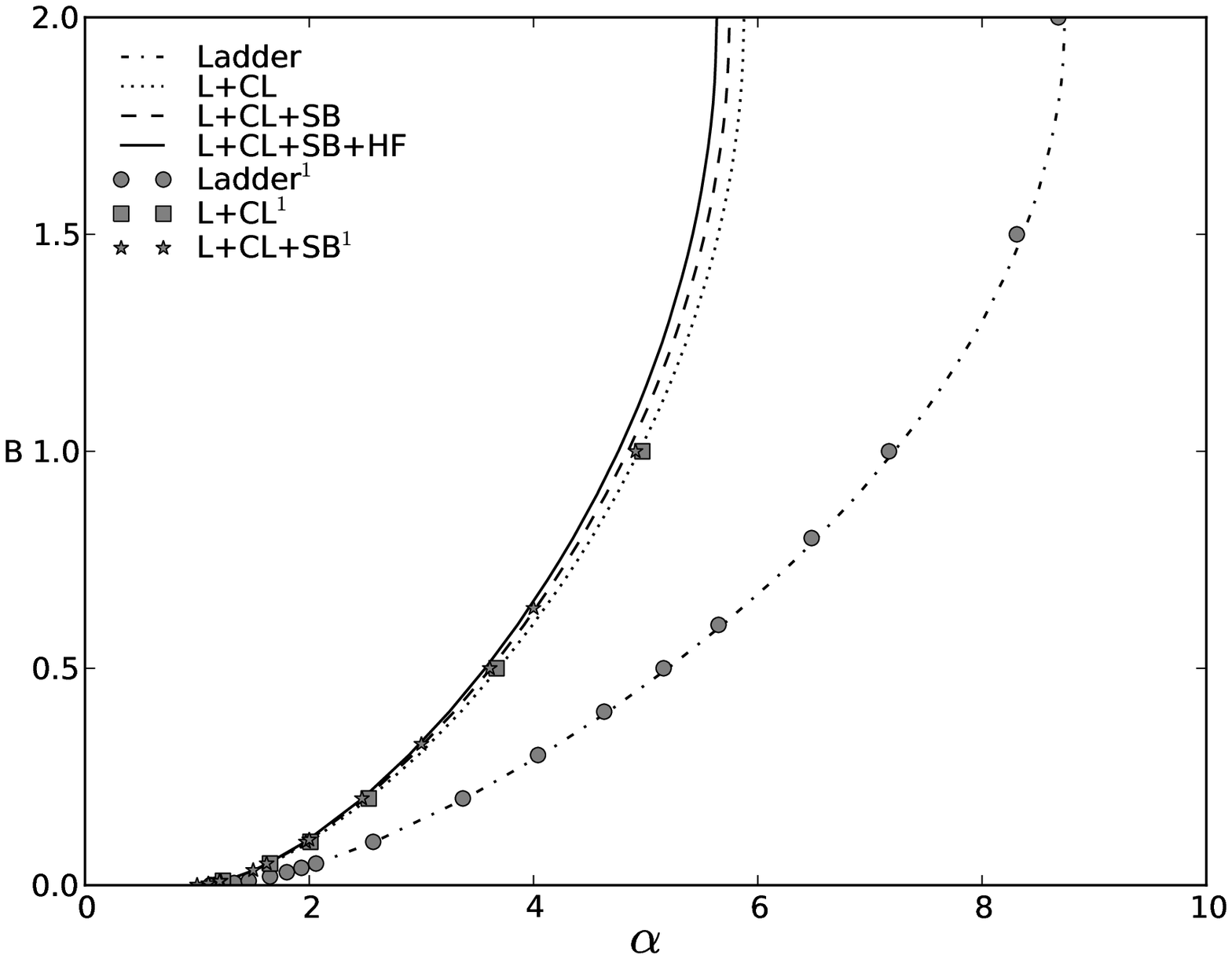}
\caption{$\beta=1$, $\lambda=0.5$. The ladder is compared with the corresponding analytic results from Eq.~(\ref{pioveralphaeqlmassnonzerolambda}). The ladder, $L+CL$, and $L+CL+SB$ are compared with the previous data. $^1$(Data from J.\ Carbonell and V.A.\ Karmanov \cite{Carbonell:2006zz} and personal communication with J.\ Carbonell \cite{carbonellprivate}.)}
\label{comparekarmabeta1lambda05}
\end{center}
\end{minipage}
\end{figure}

In Fig.~\ref{beta1lambda25}, we consider the case of $\beta=1, \lambda=2.5$ where we find that the particle-antiparticle creation/annihilation contribution from $V^{HF}$ becomes quite substantial. The contribution of $V^{HF}$ is much greater than that of $V^{SB}$, with or without the self-energy corrections. In particular, with the self-energy corrections, the $V^{HF}$ contribution becomes large enough to make the curve $L+CL+SB+HF+${\it self} go above the ladder curve. We also note that the bound states can be formed only with a sufficiently large coupling since it is difficult to exchange a large mass particle with insufficient coupling strength. Moreover, due to the short range interaction in this case, the effects of the self-energy and cross-ladder seem to appear quite different when compared to those in the case of $\lambda=0$. Although we did not include the whole curve of $L+${\it self}, we note that $\alpha$ reaches $57.4$ when the binding energy $B$ becomes $2m$. 

In Fig.~\ref{comparekarmabeta1lambda05}, the ladder (thin black dot-dashed), $L+CL$ (thin black dotted), and $L+CL+SB$ (thin black dashed) for the case of $\beta=1, \lambda=0.5$ are compared with the corresponding numerical data (solid circles, squares, and stars, respectively) given by J.\ Carbonell and V.A.\ Karmanov \cite{Carbonell:2006zz,carbonellprivate}. Our numerical results agree well with their results. This indicates that our variational wavefunction in Eq.~(\ref{psi}), based on the solution for the Hulthen potential, is reasonable for $\lambda\not=0$ case. The thin black solid line for $L+CL+SB+HF$ is also displayed in this figure to show the $V^{HF}$ effect.

\begin{figure}
\begin{minipage}{0.5\textwidth}
\begin{center}
\includegraphics[width=1.0\linewidth]{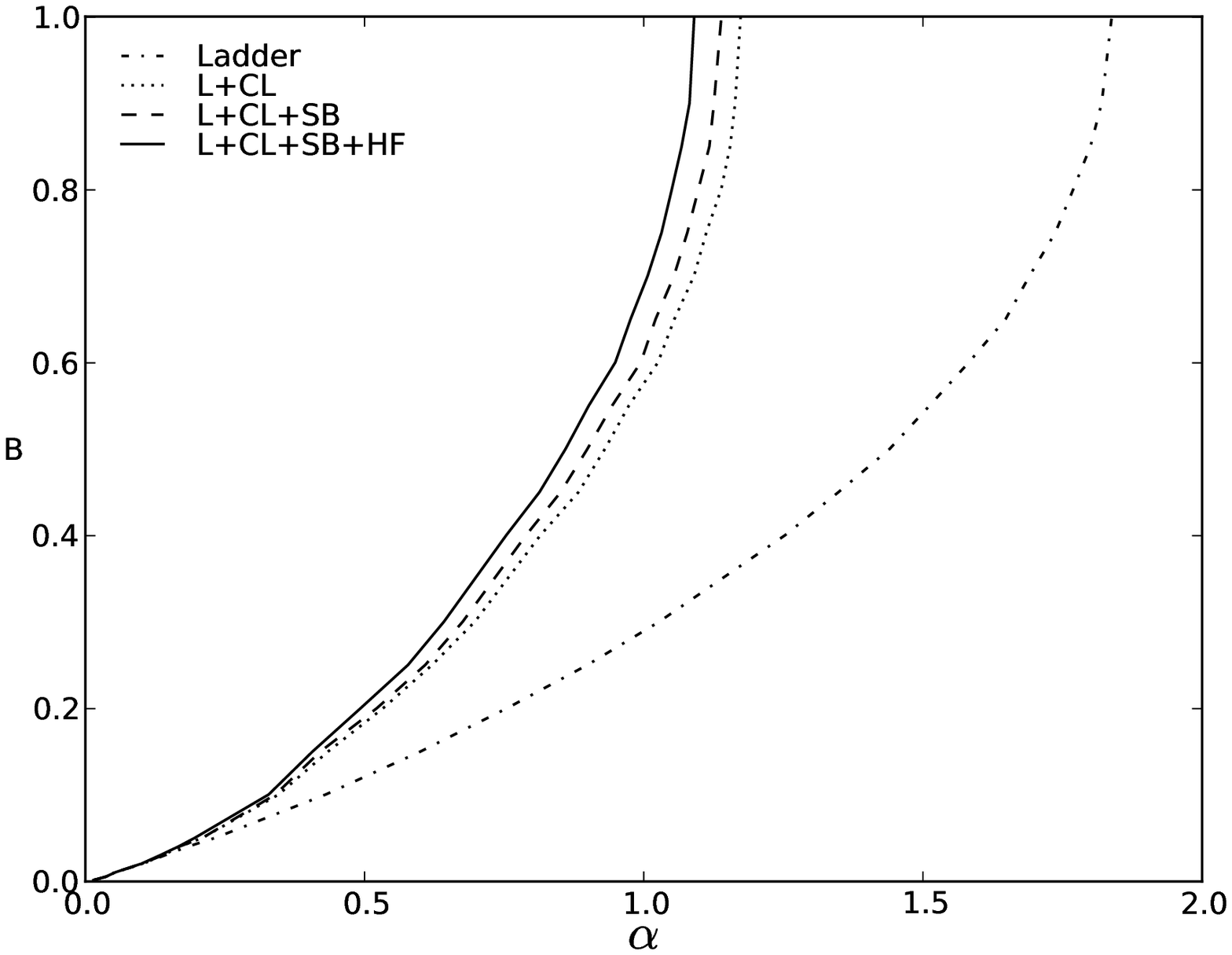}
\caption{$\beta=5.446\times 10^{-4}$, $\lambda=0$. The numerical results of the ladder, $L+CL$, $L+CL+SB$, and $L+CL+SB+HF$ without the self-energy corrections obtained by the variational principle.}
\label{beta0lambda0}
\end{center}
\end{minipage}
\hspace{0.0\linewidth}
\begin{minipage}{0.5\textwidth}
\begin{center}
\includegraphics[width=1.0\linewidth]{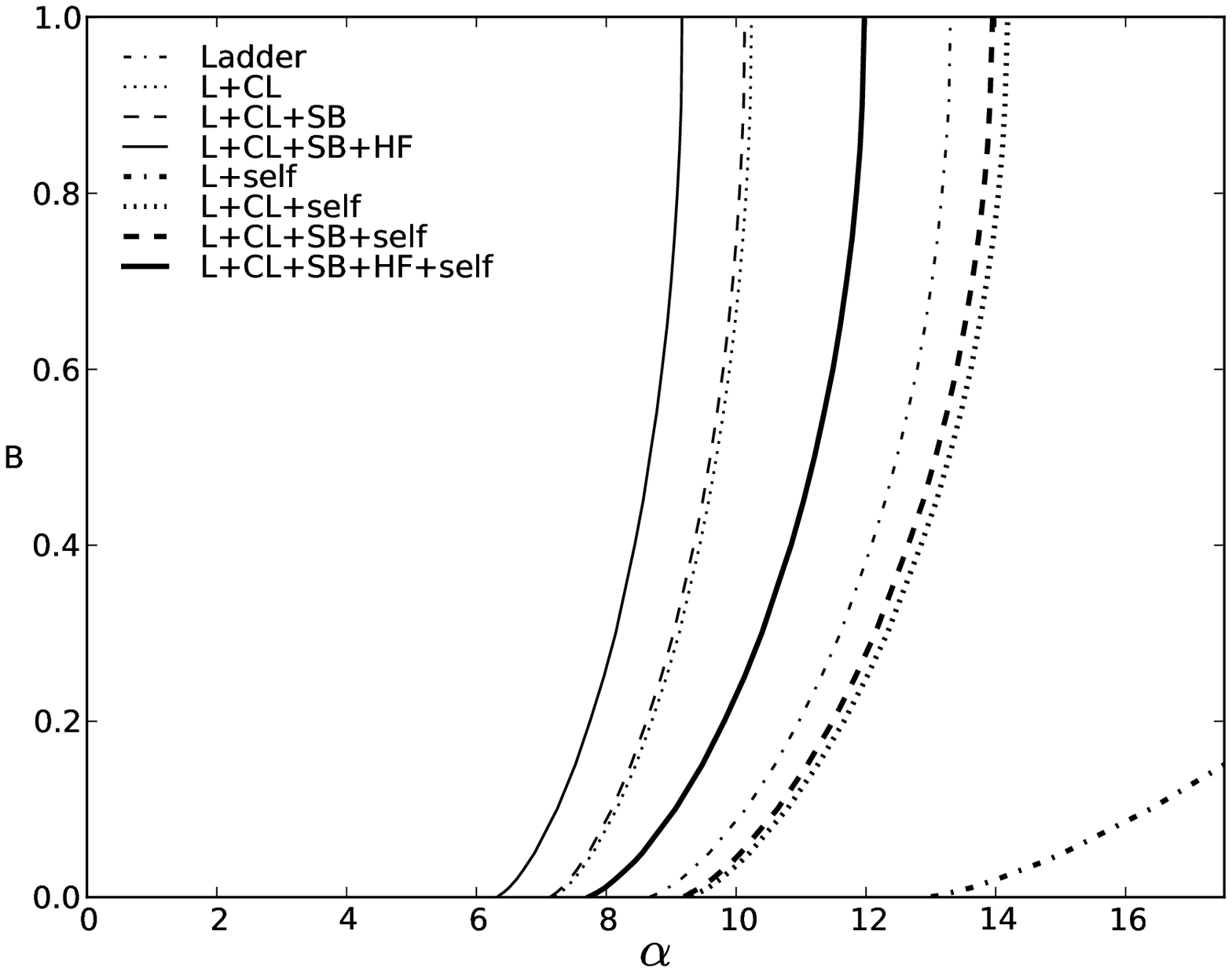}
\caption{$\beta=5.446\times 10^{-4}$, $\lambda=2.5$. The numerical results of the ladder, $L+CL$, $L+CL+SB$, and $L+CL+SB+HF$ with and without the self-energy corrections obtained by the variational principle.}
\label{beta0lambda25}
\end{center}
\end{minipage}
\end{figure}

Fig.~\ref{beta0lambda0} presents the unequal mass case of $\beta=5.446\times 10^{-4}$, $\lambda=0$. The ladder (thin black dot-dashed), $L+CL$ (thin black dotted), and $L+CL+SB$ (thin black dashed), and $L+CL+SB+HF$ (thin black solid line) are shown in this figure. Here, we see that the particle-antiparticle creation/annihilation contribution from $V^{HF}$ is non-negligible although the exchange particle mass is zero ($\lambda=0$). All the results including the self-energy effects, such as $L+${\it self}, $L+CL+${\it self}, $L+CL+SB+${\it self}, and $L+CL+SB+HF+${\it self}, are not shown in the figure because they are all strongly suppressed in this case.  

Fig.~\ref{beta0lambda25} shows the case of $\beta=5.446\times 10^{-4}$, $\lambda=2.5$. We find that the particle-antiparticle creation/annihilation contribution from $V^{HF}$ is enhanced most significantly in this case. We also note that the attractive effect of the cross-ladder contribution becomes significant when the self-energy is included as one can see from the $L+CL+${\it self}, $L+CL+SB+${\it self}, and $L+CL+SB+HF+${\it self} curves. However, the $L+${\it self} result is away from the other results and we note that $\alpha$ reaches $25.45$ as the binding energy $B$ approaches to $m$.

These results confirm that the self-energy effect is repulsive while the effect of the cross-ladder contribution is attractive. Comparing the results with and without the self-energy effect, we notice that the attractive effect from the cross-ladder contribution is more significant when the self-energy is included. Although the effects of the stretched-box and the particle-antiparticle creation/annihilation are also attractive, their contributions are not as significant as the cross-ladder contribution. In particular, the effect of the stretched-box is further reduced when $\beta\not=1$ and $\lambda\not=0$.

Although the particle-antiparticle creation/annihilation contribution is not so significant when $\beta=1$ and $\lambda=0$, it becomes non-negligible as $\beta\not=1$ and/or $\lambda\not=0$. We find that the $V^{HF}$ effect is the most significant in the case of $\beta\not=1, \lambda\not=0$ and also non-negligible for the $\beta=1, \lambda\not=0$ and $\beta\not=1, \lambda=0$ cases. 

\subsection{Zero Binding Limit} 
\begin{figure}
\begin{center}
\includegraphics[scale=0.62]{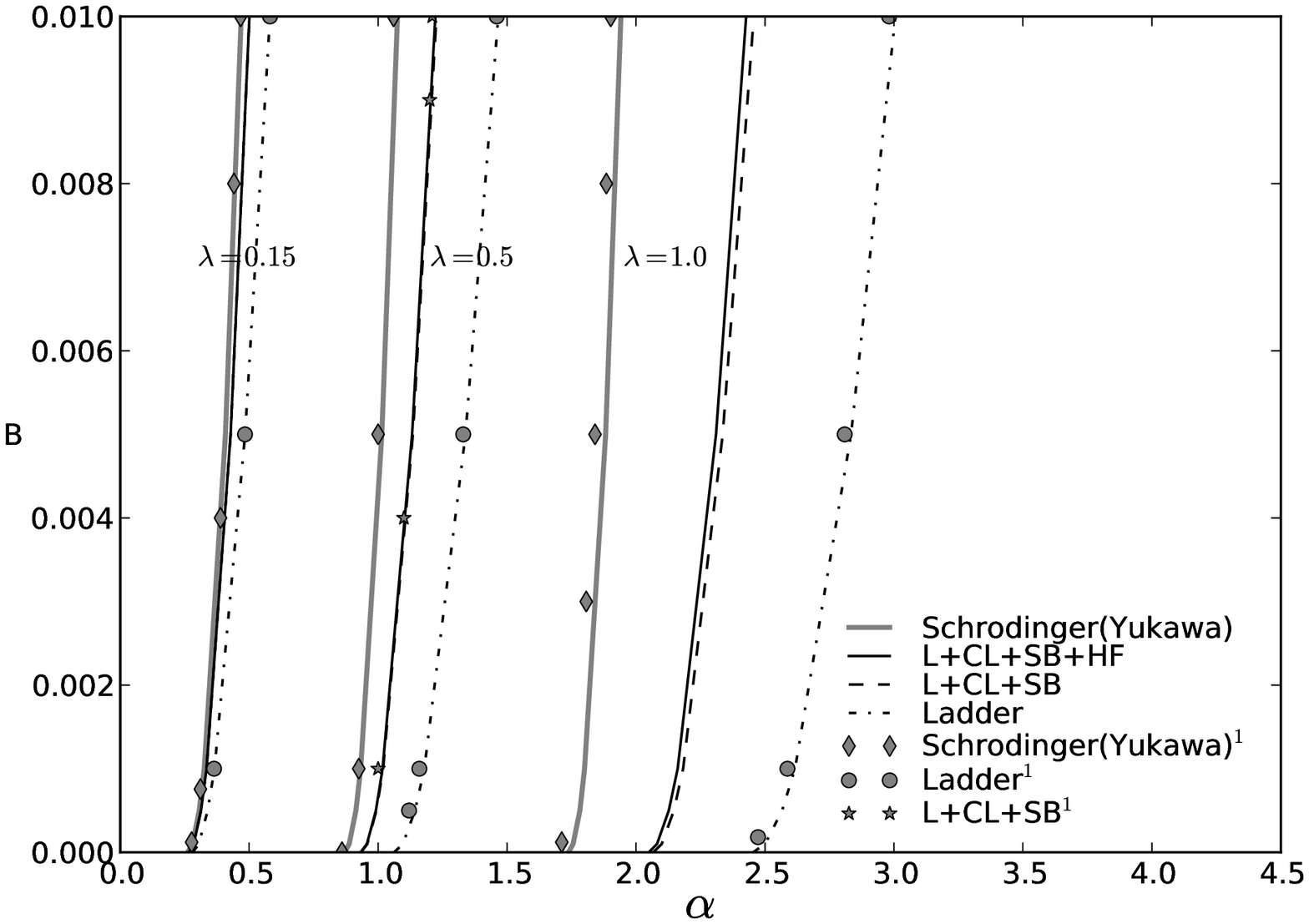}
\caption{Zero binding energy region of the ladder, LD+CL+SB, and LD+CL+SB+HF results compared with the non-relativistic ones (Schr\"{o}dinger equation with Yukawa potential). $\beta=1$, $\lambda=0.15, 0.5$, and $1.0$. $^1$(M.\ Mangin-Brinet and J.\ Carbonell \cite{ManginBrinet:1999fh}.)} 
\label{zerobindinglimitwithkarma0150510version2}
\end{center}
\end{figure}

\begin{figure}
\begin{center}
\includegraphics[scale=0.62]{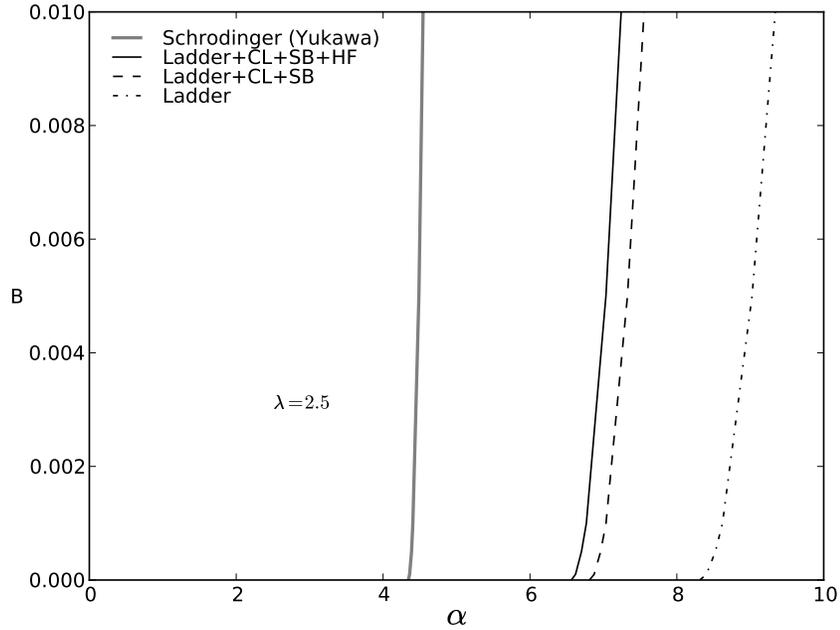}
\caption{Zero binding energy region of the ladder, LD+CL+SB and LD+CL+SB+HF results compared to the non-relativistic ones (Schr\"{o}dinger equation with Yukawa potential). $\beta=1$, $\lambda=2.5$.}
\label{zerobindinglimit25}
\end{center}
\end{figure}

It has been indicated in \cite{Carbonell:2006zz,ManginBrinet:1999fh,ManginBrinet:2000hq,ManginBrinet:2001ma} that for the non-zero mass of the exchange particle $\chi$, the relativistic ladder results in the weak binding limit $B \to0$ do not coincide with the non-relativistic result. We confirmed this result as shown in Fig.~\ref{zerobindinglimitwithkarma0150510version2}. Our results of ladder (thin black dot-dashed), $L+CL+SB$ (thin black dashed), and $L+CL+SB+HF$ (thin black solid line) are depicted for $\lambda=0.15, 0.5$, and $1.0$, and compared with the available ladder (solid circles) and $L+CL+SB$ (stars) results from Ref.~\cite{ManginBrinet:1999fh}. Also our results of the Schr\"{o}dinger equation with the Yukawa potential obtained by the variational principle (thick grey solid line) are compared with the corresponding results (diamonds) from Ref.~\cite{ManginBrinet:1999fh}. Our results are in close agreement with all of the available results from Ref.~\cite{ManginBrinet:1999fh}. As one can see in Fig.~\ref{zerobindinglimitwithkarma0150510version2}, the discrepancy between the non-relativistic limit and the zero binding limit becomes larger as the exchange particle mass increases. However, the cross-ladder and stretched-box reduce the discrepancy. We notice that the particle-antiparticle creation/annihilation contribution from $V^{HF}$ further reduces the discrepancy although it is not as significant as the cross-ladder and stretched-box. To show the more noticeable $V^{HF}$ effect, we present the $\lambda=2.5$ case in Fig.~\ref{zerobindinglimit25}. One may expect that the discrepancy would not be completely removed unless all the irreducible kernels are included in the relativistic bound-state equation.

\subsection{Wavefunction Renormalization and Probabilities of Lowest and Higher Fock-States} 
In Subsection 2.4, we defined the probabilities of finding the lowest two-body Fock-state and the higher Fock-states with and without the self-energy correction. In particular, including the self-energy correction, we have renormalized the two-body wavefunction and defined the corresponding probability with the renormalized wavefunction (see Eq.~(\ref{plowwithself})). The probability to find the higher Fock-states with the self-energy effect was also defined by Eq.~(\ref{phighwithself}).

\begin{figure}
\begin{center}
\includegraphics[width=0.75\linewidth]{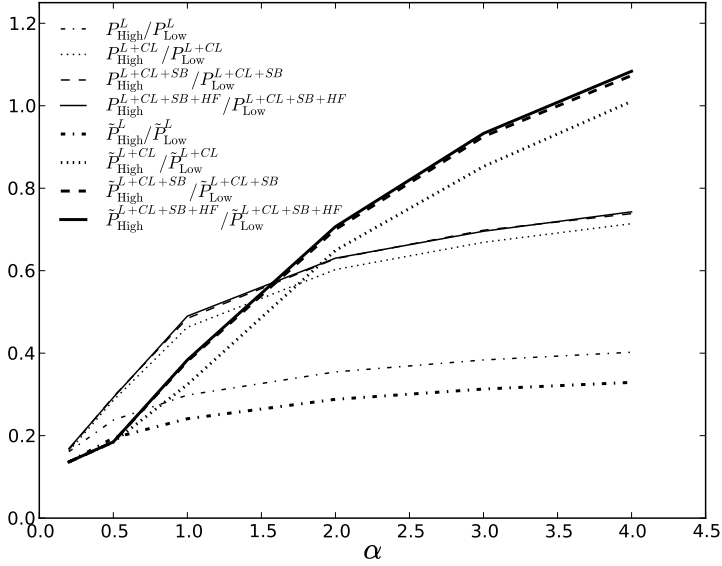}
\caption{$\beta=1$, $\lambda=0$. The numerical results of the probability ratios of $P_{\rm High}/P_{\rm Low}$ (without self-energy corrections) and $\tilde{P}_{\rm High}/\tilde{P}_{\rm Low}$ (with self-energy corrections) for the ladder, $L+CL$, $L+CL+SB$, and $L+CL+SB+HF$.}
\label{p2p3beta1lambda0}
\end{center}
\end{figure}

\begin{figure}
\begin{center}
\includegraphics[width=0.75\linewidth]{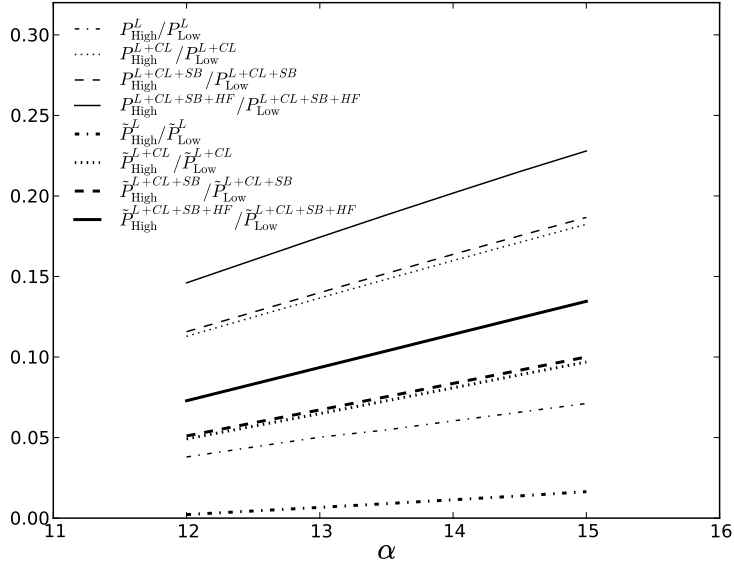}
\caption{$\beta=1$, $\lambda=2.5$. The numerical results of the probability ratios of $P_{\rm High}/P_{\rm Low}$ (without self-energy corrections) and $\tilde{P}_{\rm High}/\tilde{P}_{\rm Low}$ (with self-energy corrections) for the ladder, $L+CL$, $L+CL+SB$, and $L+CL+SB+HF$.}
\label{p2p3beta1lambda25}
\end{center}
\end{figure}

In this subsection, we present the numerical results of the probability ratios of $P_{\rm High}/P_{\rm Low}$ without the self-energy correction and $\tilde{P}_{\rm High}/\tilde{P}_{\rm Low}$ with the self-energy correction. In Figs.~\ref{p2p3beta1lambda0} and \ref{p2p3beta1lambda25}, we show the probability ratios for $P_{\rm High}^{L}/P_{\rm Low}^{L}$ (thin black dot-dashed), $P_{\rm High}^{L+CL}/P_{\rm Low}^{L+CL}$ (thin black dotted), $P_{\rm High}^{L+CL+SB}/P_{\rm Low}^{L+CL+SB}$ (thin black dashed), and $P_{\rm High}^{L+CL+SB+HF}/P_{\rm Low}^{L+CL+SB+HF}$ (thin black solid line) as well as the probability ratios for $\tilde{P}_{\rm High}^{L}/\tilde{P}_{\rm Low}^{L}$ (thick black dot-dashed), $\tilde{P}_{\rm High}^{L+CL}/\tilde{P}_{\rm Low}^{L+CL}$ (thick black dotted), $\tilde{P}_{\rm High}^{L+CL+SB}/\tilde{P}_{\rm Low}^{L+CL+SB}$ (thick black dashed), and $\tilde{P}_{\rm High}^{L+CL+SB+HF}/\tilde{P}_{\rm Low}^{L+CL+SB+HF}$ (thick black solid line), for $\lambda=0$ and $2.5$, respectively, with $\beta=1$. 

Wether or not the self-energy corrections are included, we note that the probability ratios become larger as the coupling constant $\alpha$ increases. This indicates that the sector of the higher Fock-states becomes more significant as $\alpha$ increases. The probability ratio of the ladder with the self-energy corrections is constantly lower than that of the ladder without the self-energy corrections due to the highly repulsive effect from the self-energy term. 

In the case of $\beta=1$ and $\lambda=0$, as shown in Fig.~\ref{p2p3beta1lambda0}, the attractive effect from the cross-ladder term in $\tilde{P}_{\rm{High}}^{L+CL}/\tilde{P}_{\rm{Low}}^{L+CL}$ becomes so large that $\tilde{P}_{\rm{High}}^{L+CL}/\tilde{P}_{\rm{Low}}^{L+CL}$ even crosses over $P_{\rm{High}}^{L+CL}/P_{\rm{Low}}^{L+CL}$ around $\alpha=1.5$. This indicates that the attractive effect from the cross-ladder kernel is more enhanced when the self-energy corrections are included. This behavior of the enhanced attractive effect from the cross-ladder contribution with the inclusion of the self-energy correction is consistent with what we have observed in our spectrum calculation (see e.g. Fig.~\ref{beta1lambda0includeeverything} and the corresponding text in Subsection 3.1). This consistency between the results of the wavefunction related observables and the mass spectra manifests the correlation between the wavefunction and the corresponding mass spectrum as the eigenfunction and the corresponding eigenvalue must be correlated in the bound-state calculation. 

Due to the highly repulsive effect from the self-energy, the probability ratio $\tilde{P}_{\rm High}^{L}/\tilde{P}_{\rm Low}^{L}$ (with self-energy) is lower than $P_{\rm High}^{L}/P_{\rm Low}^{L}$ (without self-energy) indicating that the probability of finding the higher Fock-states is lower when the self-energy effect is included without any significant attractive channel such as the cross-ladder contribution. When the cross-ladder contribution is included, however, there is a dramatic increase in the probability of finding the higher Fock-states. Without the self-energy effect, the ratio $P_{\rm{High}}^{L+CL}/P_{\rm{Low}}^{L+CL}$ is much higher than the ratio $P_{\rm{High}}^{L}/P_{\rm{Low}}^{L}$. With the self-energy effect, the increment of the ratio $\tilde{P}_{\rm High}^{L+CL}/\tilde{P}_{\rm Low}^{L+CL}$ is more dramatic than that of the ratio $\tilde{P}_{\rm High}^{L}/\tilde{P}_{\rm Low}^{L}$ as the coupling constant $\alpha$ becomes larger, even crossing over $P_{\rm{High}}^{L+CL}/P_{\rm{Low}}^{L+CL}$ around $\alpha=1.5$.

In the case of $\beta=1$ and $\lambda=2.5$, as shown in Fig.~\ref{p2p3beta1lambda25}, we see again that the effect of the cross-ladder contribution is highly attractive and enhances the probability of finding the higher Fock-states wether the self-energy corrections are included or not. However, the increment of the ratio $\tilde{P}_{\rm High}^{L+CL}/\tilde{P}_{\rm Low}^{L+CL}$ is not as dramatic as in the case of $\lambda=0$ and the ratio $\tilde{P}_{\rm High}^{L+CL}/\tilde{P}_{\rm Low}^{L+CL}$ does not cross over the ratio $P_{\rm High}^{L+CL}/P_{\rm Low}^{L+CL}$. This appears to be consistent with the reasoning that the probability of finding the higher Fock-states, including the very massive exchange particle of $\lambda=2.5$, is much less than the corresponding probability involving the massless exchange particle of $\lambda=0$.

\section{Conclusion} 
We have developed a variational method for solving the scalar field model theoriy ($m_1\not=m_2,  \lambda\not=0$) up to second order in the coupling constant $\alpha$. The solutions for the Schr\"{o}dinger equation with the Hulthen potential can be taken as a reasonable variational wavefunction for the $\lambda\not=0$ case. We go beyond the light-front ladder approximation and extend to the full LFD kernel including the ladder, cross-ladder, stretched-box, and particle-antiparticle creation/annihilation contribution ($V^{HF}$). We also obtain the light-front two-body equation including the self-energy corrections in the case of $m_1\not=m_2,  \lambda\not=0$. Our numerical results of the light-front formalism agree well with the previous numerical data \cite{ManginBrinet:1999fh,Carbonell:2006zz}. They are also consistent with our analytic result in the weak binding limit. 

Our results between the coupling constant $\alpha$ and the binding energy $B$ indicate that the self-energy effect is repulsive while all other effects that we took into account, such as the ladder, cross-ladder, stretched-box, and particle-antiparticle creation/annihilation contributions, are attractive. Besides the ladder, the cross-ladder effect is most significant  among the attractive contributions. The effect of the cross-ladder is more significant when the self-energy is included. 

Although the particle-antiparticle creation/annihilation contribution from $V^{HF}$ is not so significant for $\beta=1$ and $\lambda=0$, it becomes non-negligible as $\beta\not=1$ and/or $\lambda\not=0$. In particular, we find that the $V^{HF}$ effect is the most significant in the case of $\beta\not=1, \lambda\not=0$ and also non-negligible in the $\beta=1, \lambda\not=0$ or $\beta\not=1, \lambda=0$ case. In contrast to this behavior of the $V^{HF}$ contribution, the effect of the stretched-box $V^{SB}$ is more reduced as $\beta\not=1$ and/or $\lambda\not=0$.

We also note that the spectrum of the bound-state is intimately correlated with the corresponding bound-state wavefunction. We see that the attractive effect of the cross-ladder enhances the probability of finding the higher Fock-states whether the self-energy corrections are included or not. In the case of $\lambda=0$, without the self-energy effect, the ratio $P_{\rm{High}}^{L+CL}/P_{\rm{Low}}^{L+CL}$ is much higher than the ratio $P_{\rm{High}}^{L}/P_{\rm{Low}}^{L}$. Including the self-energy effect, the increment of the probability ratio $\tilde{P}_{\rm High}^{L+CL}/\tilde{P}_{\rm Low}^{L+CL}$ is more dramatic than that of the ratio $\tilde{P}_{\rm High}^{L}/\tilde{P}_{\rm Low}^{L}$ as $\alpha$ increases. This seems consistent with the result of the spectrum calculation exhibiting the more pronounced cross-ladder effect when the self-energy is included. As the exchange particle mass increases, however, the probability of finding the higher Fock-states may be reduced as one can understand from the derivative of energy denominator involving the exchange particle mass. As such, in the case of $\lambda=2.5$, the increment of the ratio $\tilde{P}_{\rm High}^{L+CL}/\tilde{P}_{\rm Low}^{L+CL}$ is not as dramatic as in the case of $\lambda=0$. The probability of finding the higher Fock-states with the massive exchange particle of $\lambda=2.5$ is much less than the corresponding probability involving the massless exchange particle of $\lambda=0$.

\section*{Acknowledgements}

We thank Jaume Carbonell for providing the numerical data that we compare with our results in this work. We would also like to thank Ben Bakker, Franz Gross, and Alfredo Suzuki for their interest and valuable comments on this work. One of us (Y.T.) acknowledges useful discussions with Shinya Matsuzaki. This work is supported by the U.S. Department of energy (No. DE-FG02-03ER41260).



\appendix{}
\numberwithin{equation}{section}
\begin{center}
\section{\label{appendixA}Full Expression of the Kernel}
\end{center}
Corresponding to Figs.~\ref{stretchfig}, \ref{crossfig}, and \ref{hffig}, respectively, the stretched-box ($V^{SB}$), cross-ladder ($V^{CL}$), and higher-Fock ($V^{HF}$) kernels are written in the form:
\begin{align}
&V^{SB}=\frac{1}{z(x-z)(z-y)(1-z)}\sum_{i=1,2}V_i\,,\\
&V^{CL}=\frac{1}{z(x-z)(z-y)(1-x-y+z)}\sum_{i=3}^{8}V_i\,,\\
&V^{HF}=\frac{1}{z(x-z)(z-y)(1-x-y+z)}\left[V_A+V_B\right]\,,
\end{align}
where $V_1,V_2,V_3,V_4,V_5,V_6,V_7,V_8,V_A$, and $V_B$ are given by
\begin{align}
V_1=&(M^2 - \frac{\vec{k}_{\perp}^2 + m_1^2}{x} - \frac{(\vec{j}_{\perp}-\vec{k}_{\perp})^2 + \lambda^2}{z-x} - \frac{\vec{j}_{\perp}^2 + m_2^2}{1-z})^{-1}\nonumber\\
\times&(M^2 - \frac{\vec{k}_{\perp}^2 + m_1^2}{x}- \frac{(\vec{j}_{\perp}-\vec{k}_{\perp})^2 + \lambda^2}{z-x}- \frac{(\vec{l}_{\perp}-\vec{j}_{\perp})^2 + \lambda^2}{y-z}- \frac{\vec{l}_{\perp}^2 + m_2^2}{1-y})^{-1}\nonumber\\
\times&(M^2 - \frac{\vec{j}_{\perp}^2 + m_1^2}{z} - \frac{(\vec{l}_{\perp}-\vec{j}_{\perp})^2 + \lambda^2}{y-z}- \frac{\vec{l}_{\perp}^2 + m_2^2}{1-y})^{-1}\,,
\end{align}
\begin{align}
V_2=&(M^2 - \frac{\vec{j}_{\perp}^2 + m_1^2}{z} - \frac{(\vec{k}_{\perp}-\vec{j}_{\perp})^2 + \lambda^2}{x-z} - \frac{\vec{k}_{\perp}^2 + m_2^2}{1-x})^{-1}\nonumber\\
\times&(M^2 - \frac{\vec{l}_{\perp}^2 + m_1^2}{y}- \frac{(\vec{j}_{\perp}-\vec{l}_{\perp})^2 + \lambda^2}{z-y}- \frac{\vec{j}_{\perp}^2 + m_2^2}{1-z})^{-1}\nonumber\\
\times&(M^2 - \frac{\vec{l}_{\perp}^2 + m_1^2}{y}-\frac{(\vec{k}_{\perp}-\vec{j}_{\perp})^2 + \lambda^2}{x-z}-\frac{(\vec{j}_{\perp}-\vec{l}_{\perp})^2 + \lambda^2}{z-y}- \frac{\vec{k}_{\perp}^2 + m_2^2}{1-x})^{-1}\,,
\end{align}
\begin{align}
V_3=-&(M^2 - \frac{\vec{j}_{\perp}^2 + m_1^2}{z} - \frac{(\vec{l}_{\perp}-\vec{j}_{\perp})^2 + \lambda^2}{y-z} - \frac{\vec{l}_{\perp}^2 + m_2^2}{1-y})^{-1}\nonumber\\
\times&(M^2 - \frac{\vec{j}_{\perp}^2 + m_1^2}{z}- \frac{(\vec{l}_{\perp}-\vec{j}_{\perp})^2 + \lambda^2}{y-z}- \frac{(\vec{k}_{\perp}-\vec{j}_{\perp})^2 + \lambda^2}{x-z}- \frac{(\vec{j}_{\perp}-\vec{k}_{\perp}-\vec{l}_{\perp})^2 +m_2^2}{1-x-y+z})^{-1}\nonumber\\
\times&(M^2 - \frac{\vec{j}_{\perp}^2 + m_1^2}{z} - \frac{(\vec{k}_{\perp}-\vec{j}_{\perp})^2 + \lambda^2}{x-z}- \frac{\vec{k}_{\perp}^2 + m_2^2}{1-x})^{-1}\,,
\end{align}
\begin{align}
V_4=-&(M^2 - \frac{\vec{k}_{\perp}^2 + m_1^2}{x}-\frac{(\vec{l}_{\perp}-\vec{j}_{\perp})^2 + \lambda^2}{y-z}-\frac{(\vec{j}_{\perp}-\vec{k}_{\perp}-\vec{l}_{\perp})^2 +m_2^2}{1-x-y+z})^{-1}\nonumber\\
\times&(M^2 - \frac{\vec{j}_{\perp}^2 + m_1^2}{z}- \frac{(\vec{k}_{\perp}-\vec{j}_{\perp})^2 + \lambda^2}{x-z}- \frac{(\vec{l}_{\perp}-\vec{j}_{\perp})^2 + \lambda^2}{y-z}- \frac{(\vec{j}_{\perp}-\vec{k}_{\perp}-\vec{l}_{\perp})^2 +m_2^2}{1-x-y+z})^{-1}\nonumber\\
\times&(M^2 - \frac{\vec{l}_{\perp}^2 + m_1^2}{y} - \frac{(\vec{k}_{\perp}-\vec{j}_{\perp})^2 + \lambda^2}{x-z}- \frac{(\vec{j}_{\perp}-\vec{k}_{\perp}-\vec{l}_{\perp})^2 +m_2^2}{1-x-y+z})^{-1}\,,
\end{align}
\begin{align}
V_5=-&(M^2 - \frac{\vec{k}_{\perp}^2 + m_1^2}{x} - \frac{(\vec{l}_{\perp}-\vec{j}_{\perp})^2 + \lambda^2}{y-z}- \frac{(\vec{j}_{\perp}-\vec{k}_{\perp}-\vec{l}_{\perp})^2 +m_2^2}{1-x-y+z})^{-1}\nonumber\\
\times&(M^2 - \frac{\vec{j}_{\perp}^2 + m_1^2}{z}- \frac{(\vec{k}_{\perp}-\vec{j}_{\perp})^2 + \lambda^2}{x-z} - \frac{(\vec{l}_{\perp}-\vec{j}_{\perp})^2 + \lambda^2}{y-z}- \frac{(\vec{j}_{\perp}-\vec{k}_{\perp}-\vec{l}_{\perp})^2 +m_2^2}{1-x-y+z})^{-1}\nonumber\\
\times&(M^2 - \frac{\vec{j}_{\perp}^2 + m_1^2}{z} - \frac{(\vec{l}_{\perp}-\vec{j}_{\perp})^2 + \lambda^2}{y-z} - \frac{\vec{l}_{\perp}^2 + m_2^2}{1-y})^{-1}\,,
\end{align}
\begin{align}
V_6=-&(M^2 - \frac{\vec{j}_{\perp}^2 + m_1^2}{z} - \frac{(\vec{k}_{\perp}-\vec{j}_{\perp})^2 + \lambda^2}{x-z}- \frac{\vec{k}_{\perp}^2 + m_2^2}{1-x})^{-1}\nonumber\\
\times&(M^2 - \frac{\vec{j}_{\perp}^2 + m_1^2}{z}- \frac{(\vec{k}_{\perp}-\vec{j}_{\perp})^2 + \lambda^2}{x-z}- \frac{(\vec{l}_{\perp}-\vec{j}_{\perp})^2 + \lambda^2}{y-z}- \frac{(\vec{j}_{\perp}-\vec{k}_{\perp}-\vec{k}_{\perp})^2 +m_2^2}{1-x-y+z})^{-1}\nonumber\\
\times&(M^2 - \frac{\vec{l}_{\perp}^2 + m_1^2}{y} - \frac{(\vec{k}_{\perp}-\vec{j}_{\perp})^2 + \lambda^2}{x-z}- \frac{(\vec{j}_{\perp}-\vec{k}_{\perp}-\vec{l}_{\perp})^2 +m_2^2}{1-x-y+z})^{-1}\,,
\end{align}
\begin{align}
V_7=&(M^2 - \frac{\vec{k}_{\perp}^2 + m_1^2}{x} - \frac{(\vec{l}_{\perp}-\vec{j}_{\perp})^2 + \lambda^2}{y-z}- \frac{(\vec{j}_{\perp}-\vec{k}_{\perp}-\vec{l}_{\perp})^2 +m_2^2}{1-x-y+z})^{-1}\nonumber\\
\times&(M^2 - \frac{\vec{k}_{\perp}^2 + m_1^2}{x}- \frac{(\vec{j}_{\perp}-\vec{k}_{\perp})^2 + \lambda^2}{z-x}- \frac{(\vec{l}_{\perp}-\vec{j}_{\perp})^2 + \lambda^2}{y-z}- \frac{\vec{l}_{\perp}^2 +m_2^2}{1-y})^{-1}\nonumber\\
\times&(M^2 - \frac{\vec{j}_{\perp}^2 + m_1^2}{z} - \frac{(\vec{l}_{\perp}-\vec{j}_{\perp})^2 + \lambda^2}{y-z}- \frac{\vec{l}_{\perp}^2 + m_2^2}{1-y})^{-1}\,,
\end{align}
\begin{align}
V_8=&(M^2 - \frac{\vec{j}_{\perp}^2 + m_1^2}{z} - \frac{(\vec{k}_{\perp}-\vec{j}_{\perp})^2 + \lambda^2}{x-z} - \frac{\vec{k}_{\perp}^2 + m_2^2}{1-x})^{-1}\nonumber\\
\times&(M^2 - \frac{\vec{l}_{\perp}^2 + m_1^2}{y}- \frac{(\vec{k}_{\perp}-\vec{j}_{\perp})^2 + \lambda^2}{x-z}- \frac{(\vec{j}_{\perp}-\vec{l}_{\perp})^2 + \lambda^2}{z-y}- \frac{\vec{k}_{\perp}^2 +m_2^2}{1-x})^{-1}\nonumber\\
\times&(M^2 - \frac{\vec{l}_{\perp}^2 + m_1^2}{y} - \frac{(\vec{k}_{\perp}-\vec{j}_{\perp})^2 + \lambda^2}{x-z}- \frac{(\vec{j}_{\perp}-\vec{k}_{\perp}-\vec{l}_{\perp})^2 +m_2^2}{1-x-y+z})^{-1}\,,
\end{align}
\begin{align}
V_A=&(M^2 - \frac{\vec{l}_{\perp}^2 + m_1^2}{y} - \frac{(\vec{k}_{\perp}-\vec{j}_{\perp})^2 + \lambda^2}{x-z}- \frac{(\vec{j}_{\perp}-\vec{k}_{\perp}-\vec{l}_{\perp})^2 +m_2^2}{1-x-y+z})^{-1}\nonumber\\
\times&(M^2 - \frac{\vec{k}_{\perp}^2 + m_1^2}{x}- \frac{\vec{j}_{\perp}^2 + m_1^2}{z}- \frac{\vec{l}_{\perp}^2 + m_1^2}{y}-\frac{(\vec{j}_{\perp}-\vec{k}_{\perp}-\vec{l}_{\perp})^2 +m_2^2}{1-x-y+z})^{-1}\nonumber\\
\times&(M^2 - \frac{\vec{k}_{\perp}^2 + m_1^2}{x} - \frac{(\vec{l}_{\perp}-\vec{j}_{\perp})^2 + \lambda^2}{y-z}-\frac{(\vec{j}_{\perp}-\vec{k}_{\perp}-\vec{l}_{\perp})^2 +m_2^2}{1-x-y+z})^{-1}\,,
\end{align}
\begin{align}
V_B=&(M^2 - \frac{\vec{j}_{\perp}^2 + m_1^2}{z} - \frac{(\vec{k}_{\perp}-\vec{j}_{\perp})^2 + \lambda^2}{x-z} - \frac{\vec{k}_{\perp}^2 + m_2^2}{1-x})^{-1}\nonumber\\
\times&(M^2 - \frac{\vec{j}_{\perp}^2 + m_1^2}{z}- \frac{\vec{k}_{\perp}^2 + m_2^2}{1-x}- \frac{\vec{l}_{\perp}^2 + m_2^2}{1-y}- \frac{(\vec{j}_{\perp}-\vec{k}_{\perp}-\vec{l}_{\perp})^2 +m_2^2}{1-x-y+z})^{-1}\nonumber\\
\times&(M^2 - \frac{\vec{j}_{\perp}^2 + m_1^2}{z} - \frac{(\vec{l}_{\perp}-\vec{j}_{\perp})^2 + \lambda^2}{y-z} - \frac{\vec{l}_{\perp}^2 + m_2^2}{1-y})^{-1}\,.
\end{align}

\bigskip
\bigskip

%





{\centering\section{\label{appendixb}Full Expressions of $P_{\rm Low}$ and $P_{\rm High}$}}

The full expressions of $P_{\rm Low}$ and $P_{\rm High}$ without the self-energy can be given by the following integrations:
\begin{equation}
P_{\rm Low}^L=\int \frac{dx}{x(1-x)}\frac{d^2\vec{k}_{\perp}}{16\pi^3}|\psi_L(x,\vec{k}_{\perp})|^2\,,
\end{equation}
\begin{equation}
P_{\rm Low}^{L+CL}=\int \frac{dx}{x(1-x)}\frac{d^2\vec{k}_{\perp}}{16\pi^3}|\psi_{L+CL}(x,\vec{k}_{\perp})|^2\,,
\end{equation}
\begin{equation}
P_{\rm Low}^{L+CL+SB}=\int \frac{dx}{x(1-x)}\frac{d^2\vec{k}_{\perp}}{16\pi^3}|\psi_{L+CL+SB}(x,\vec{k}_{\perp})|^2\,,
\end{equation}
\begin{equation}
P_{\rm Low}^{L+CL+SB+HF}=\int \frac{dx}{x(1-x)}\frac{d^2\vec{k}_{\perp}}{16\pi^3}|\psi_{L+CL+SB+HF}(x,\vec{k}_{\perp})|^2\,,
\end{equation}
\begin{equation}
P_{\rm High}^{L}=\left\langle -\frac{\partial(\alpha V^{L})}{\partial M^2}\right\rangle=-g^2\,\int \frac{dx}{x(1-x)}\frac{d^2\vec{k}_{\perp}}{16\pi^3}\int \frac{dy}{y(1-y)}\frac{d^2\vec{l}_{\perp}}{16\pi^3}\psi^{\dagger}_L (x,\vec{k}_{\perp})\left(\frac{\partial V^{L}}{\partial M^2}\right)\psi_L(y,\vec{l}_{\perp})\,,
\end{equation}
\begin{align}
P&_{\rm High}^{L+CL}=\left\langle -\frac{\partial(\alpha V^{L}+\alpha^2 V^{CL})}{\partial M^2}\right\rangle
\nonumber\\
&=-g^2\,\int \frac{dx}{x(1-x)}\frac{d^2\vec{k}_{\perp}}{16\pi^3}\int \frac{dy}{y(1-y)}\frac{d^2\vec{l}_{\perp}}{16\pi^3}\psi^{\dagger}_{L+CL} (x,\vec{k}_{\perp})\left(\frac{\partial V^{L}}{\partial M^2}\right)\psi_{L+CL}(y,\vec{l}_{\perp})
\nonumber\\
&-g^4\,\int \frac{dx}{x(1-x)}\frac{d^2\vec{k}_{\perp}}{16\pi^3}\int \frac{dy}{y(1-y)}\frac{d^2\vec{l}_{\perp}}{16\pi^3}\int dz\frac{d^2\vec{j}_{\perp}}{16\pi^3}\psi^{\dagger}_{L+CL} (x,\vec{k}_{\perp})\left(\frac{\partial V^{CL}}{\partial M^2}\right)\psi_{L+CL}(y,\vec{l}_{\perp})\,,
\end{align}
\begin{align}
P&_{\rm High}^{L+CL+SB}=\left\langle -\frac{\partial(\alpha V^{L}+\alpha^2 V^{CL}+\alpha^2 V^{SB})}{\partial M^2}\right\rangle
\nonumber\\
&=-g^2\,\int \frac{dx}{x(1-x)}\frac{d^2\vec{k}_{\perp}}{16\pi^3}\int \frac{dy}{y(1-y)}\frac{d^2\vec{l}_{\perp}}{16\pi^3}\psi^{\dagger}_{L+CL+SB} (x,\vec{k}_{\perp})\left(\frac{\partial V^{L}}{\partial M^2}\right)\psi_{L+CL+SB}(y,\vec{l}_{\perp})
\nonumber\\
&-g^4\,\int \frac{dx}{x(1-x)}\frac{d^2\vec{k}_{\perp}}{16\pi^3}\int \frac{dy}{y(1-y)}\frac{d^2\vec{l}_{\perp}}{16\pi^3}\int dz\frac{d^2\vec{j}_{\perp}}{16\pi^3}\psi^{\dagger}_{L+CL+SB} (x,\vec{k}_{\perp})\left(\frac{\partial V^{CL}}{\partial M^2}\right)\psi_{L+CL+SB}(y,\vec{l}_{\perp})
\nonumber\\
&-g^4\,\int \frac{dx}{x(1-x)}\frac{d^2\vec{k}_{\perp}}{16\pi^3}\int \frac{dy}{y(1-y)}\frac{d^2\vec{l}_{\perp}}{16\pi^3}\int dz\frac{d^2\vec{j}_{\perp}}{16\pi^3}\psi^{\dagger}_{L+CL+SB} (x,\vec{k}_{\perp})\left(\frac{\partial V^{SB}}{\partial M^2}\right)\psi_{L+CL+SB}(y,\vec{l}_{\perp})\,,
\end{align}
\begin{align}
P&_{\rm High}^{L+CL+SB+HF}=\left\langle -\frac{\partial(\alpha V^{L}+\alpha^2 V^{CL}+\alpha^2 V^{SB}+\alpha^2 V^{HF})}{\partial M^2}\right\rangle
\nonumber\\
&=-g^2\,\int \frac{dx}{x(1-x)}\frac{d^2\vec{k}_{\perp}}{16\pi^3}\int \frac{dy}{y(1-y)}\frac{d^2\vec{l}_{\perp}}{16\pi^3}\psi^{\dagger}_{L+CL+SB+HF} (x,\vec{k}_{\perp})\left(\frac{\partial V^{L}}{\partial M^2}\right)\psi_{L+CL+SB+HF}(y,\vec{l}_{\perp})
\nonumber\\
&-g^4\,\int \frac{dx}{x(1-x)}\frac{d^2\vec{k}_{\perp}}{16\pi^3}\int \frac{dy}{y(1-y)}\frac{d^2\vec{l}_{\perp}}{16\pi^3}\int dz\frac{d^2\vec{j}_{\perp}}{16\pi^3}\psi^{\dagger}_{L+CL+SB+HF} (x,\vec{k}_{\perp})\left(\frac{\partial V^{CL}}{\partial M^2}\right)\psi_{L+CL+SB+HF}(y,\vec{l}_{\perp})
\nonumber\\
&-g^4\,\int \frac{dx}{x(1-x)}\frac{d^2\vec{k}_{\perp}}{16\pi^3}\int \frac{dy}{y(1-y)}\frac{d^2\vec{l}_{\perp}}{16\pi^3}\int dz\frac{d^2\vec{j}_{\perp}}{16\pi^3}\psi^{\dagger}_{L+CL+SB+HF} (x,\vec{k}_{\perp})\left(\frac{\partial V^{SB}}{\partial M^2}\right)\psi_{L+CL+SB+HF}(y,\vec{l}_{\perp})
\nonumber\\
&-g^4\,\int \frac{dx}{x(1-x)}\frac{d^2\vec{k}_{\perp}}{16\pi^3}\int \frac{dy}{y(1-y)}\frac{d^2\vec{l}_{\perp}}{16\pi^3}\int dz\frac{d^2\vec{j}_{\perp}}{16\pi^3}\psi^{\dagger}_{L+CL+SB+HF} (x,\vec{k}_{\perp})\left(\frac{\partial V^{HF}}{\partial M^2}\right)\psi_{L+CL+SB+HF}(y,\vec{l}_{\perp})\,,
\end{align}
where $\psi(x,\vec{k}_{\perp})$ is the two-body wavefunction optimized by the variational parameter for the corresponding kernel and coupling constant $\alpha$. Likewise, the full expressions of $\tilde{P}_{\rm Low}$ and $\tilde{P}_{\rm High}$ with the self-energy can be given by the following integrations:
\begin{equation}
\tilde{P}_{\rm Low}^{L}=\int \frac{dx}{x(1-x)}\frac{d^2\vec{k}_{\perp}}{16\pi^3}\left[1-(\frac{\alpha m_1m_2}{\pi})\frac{\partial f (x,\vec{k}_{\perp})}{\partial M^2}\right]|\tilde{\psi}_{L}(x,\vec{k}_{\perp})|^2\,,
\end{equation}
\begin{equation}
\tilde{P}_{\rm Low}^{L+CL}=\int \frac{dx}{x(1-x)}\frac{d^2\vec{k}_{\perp}}{16\pi^3}\left[1-(\frac{\alpha m_1m_2}{\pi})\frac{\partial f (x,\vec{k}_{\perp})}{\partial M^2}\right]|\tilde{\psi}_{L+CL}(x,\vec{k}_{\perp})|^2\,,
\end{equation}
\begin{equation}
\tilde{P}_{\rm Low}^{L+CL+SB}=\int \frac{dx}{x(1-x)}\frac{d^2\vec{k}_{\perp}}{16\pi^3}\left[1-(\frac{\alpha m_1m_2}{\pi})\frac{\partial f (x,\vec{k}_{\perp})}{\partial M^2}\right]|\tilde{\psi}_{L+CL+SB}(x,\vec{k}_{\perp})|^2\,,
\end{equation}
\begin{equation}
\tilde{P}_{\rm Low}^{L+CL+SB+HF}=\int \frac{dx}{x(1-x)}\frac{d^2\vec{k}_{\perp}}{16\pi^3}\left[1-(\frac{\alpha m_1m_2}{\pi})\frac{\partial f (x,\vec{k}_{\perp})}{\partial M^2}\right]|\tilde{\psi}_{L+CL+SB+HF}(x,\vec{k}_{\perp})|^2\,,
\end{equation}
\begin{align}
\tilde{P}&_{\rm High}^{L}= \left\langle -\frac{\partial(\alpha V^{L})}{\partial M^2}\right\rangle
\nonumber\\
&=-g^2\,\int \frac{dx}{x(1-x)}\frac{d^2\vec{k}_{\perp}}{16\pi^3}\int \frac{dy}{y(1-y)}\frac{d^2\vec{l}_{\perp}}{16\pi^3}\tilde{\psi}^{\dagger}_{L} (x,\vec{k}_{\perp})\left(\frac{\partial V^{L}}{\partial M^2}\right)\tilde{\psi}_{L}(y,\vec{l}_{\perp})\,,
\end{align}
\begin{align}
\tilde{P}&_{\rm High}^{L+CL}=\left\langle -\frac{\partial(\alpha V^{L}+\alpha^2 V^{CL})}{\partial M^2}\right\rangle
\nonumber\\
&=-g^2\,\int \frac{dx}{x(1-x)}\frac{d^2\vec{k}_{\perp}}{16\pi^3}\int \frac{dy}{y(1-y)}\frac{d^2\vec{l}_{\perp}}{16\pi^3}\tilde{\psi}^{\dagger}_{L+CL} (x,\vec{k}_{\perp})\left(\frac{\partial V^{L}}{\partial M^2}\right)\tilde{\psi}_{L+CL}(y,\vec{l}_{\perp})
\nonumber\\
&-g^4\,\int \frac{dx}{x(1-x)}\frac{d^2\vec{k}_{\perp}}{16\pi^3}\int \frac{dy}{y(1-y)}\frac{d^2\vec{l}_{\perp}}{16\pi^3}\int dz\frac{d^2\vec{j}_{\perp}}{16\pi^3}\tilde{\psi}^{\dagger}_{L+CL} (x,\vec{k}_{\perp})\left(\frac{\partial V^{CL}}{\partial M^2}\right)\tilde{\psi}_{L+CL}(y,\vec{l}_{\perp})\,,
\end{align}
\begin{align}
\tilde{P}&_{\rm High}^{L+CL+SB}=\left\langle -\frac{\partial(\alpha V^{L}+\alpha^2 V^{CL}+\alpha^2 V^{SB})}{\partial M^2}\right\rangle
\nonumber\\
&=-g^2\,\int \frac{dx}{x(1-x)}\frac{d^2\vec{k}_{\perp}}{16\pi^3}\int \frac{dy}{y(1-y)}\frac{d^2\vec{l}_{\perp}}{16\pi^3}\tilde{\psi}^{\dagger}_{L+CL+SB} (x,\vec{k}_{\perp})\left(\frac{\partial V^{L}}{\partial M^2}\right)\tilde{\psi}_{L+CL+SB}(y,\vec{l}_{\perp})
\nonumber\\
&-g^4\,\int \frac{dx}{x(1-x)}\frac{d^2\vec{k}_{\perp}}{16\pi^3}\int \frac{dy}{y(1-y)}\frac{d^2\vec{l}_{\perp}}{16\pi^3}\int dz\frac{d^2\vec{j}_{\perp}}{16\pi^3}\tilde{\psi}^{\dagger}_{L+CL+SB} (x,\vec{k}_{\perp})\left(\frac{\partial V^{CL}}{\partial M^2}\right)\tilde{\psi}_{L+CL+SB}(y,\vec{l}_{\perp})
\nonumber\\
&-g^4\,\int \frac{dx}{x(1-x)}\frac{d^2\vec{k}_{\perp}}{16\pi^3}\int \frac{dy}{y(1-y)}\frac{d^2\vec{l}_{\perp}}{16\pi^3}\int dz\frac{d^2\vec{j}_{\perp}}{16\pi^3}\tilde{\psi}^{\dagger}_{L+CL+SB} (x,\vec{k}_{\perp})\left(\frac{\partial V^{SB}}{\partial M^2}\right)\tilde{\psi}_{L+CL+SB}(y,\vec{l}_{\perp})\,,
\end{align}
\begin{align}
\tilde{P}&_{\rm High}^{L+CL+SB+HF}=\left\langle -\frac{\partial(\alpha V^{L}+\alpha^2 V^{CL}+\alpha^2 V^{SB}+\alpha^2 V^{HF})}{\partial M^2}\right\rangle
\nonumber\\
&=-g^2\,\int \frac{dx}{x(1-x)}\frac{d^2\vec{k}_{\perp}}{16\pi^3}\int \frac{dy}{y(1-y)}\frac{d^2\vec{l}_{\perp}}{16\pi^3}\tilde{\psi}^{\dagger}_{L+CL+SB+HF} (x,\vec{k}_{\perp})\left(\frac{\partial V^{L}}{\partial M^2}\right)\tilde{\psi}_{L+CL+SB+HF}(y,\vec{l}_{\perp})
\nonumber\\
&-g^4\,\int \frac{dx}{x(1-x)}\frac{d^2\vec{k}_{\perp}}{16\pi^3}\int \frac{dy}{y(1-y)}\frac{d^2\vec{l}_{\perp}}{16\pi^3}\int dz\frac{d^2\vec{j}_{\perp}}{16\pi^3}\tilde{\psi}^{\dagger}_{L+CL+SB+HF} (x,\vec{k}_{\perp})\left(\frac{\partial V^{CL}}{\partial M^2}\right)\tilde{\psi}_{L+CL+SB+HF}(y,\vec{l}_{\perp})
\nonumber\\
&-g^4\,\int \frac{dx}{x(1-x)}\frac{d^2\vec{k}_{\perp}}{16\pi^3}\int \frac{dy}{y(1-y)}\frac{d^2\vec{l}_{\perp}}{16\pi^3}\int dz\frac{d^2\vec{j}_{\perp}}{16\pi^3}\tilde{\psi}^{\dagger}_{L+CL+SB+HF} (x,\vec{k}_{\perp})\left(\frac{\partial V^{SB}}{\partial M^2}\right)\tilde{\psi}_{L+CL+SB+HF}(y,\vec{l}_{\perp})
\nonumber\\
&-g^4\,\int \frac{dx}{x(1-x)}\frac{d^2\vec{k}_{\perp}}{16\pi^3}\int \frac{dy}{y(1-y)}\frac{d^2\vec{l}_{\perp}}{16\pi^3}\int dz\frac{d^2\vec{j}_{\perp}}{16\pi^3}\tilde{\psi}^{\dagger}_{L+CL+SB+HF} (x,\vec{k}_{\perp})\left(\frac{\partial V^{HF}}{\partial M^2}\right)\tilde{\psi}_{L+CL+SB+HF}(y,\vec{l}_{\perp})\,.
\end{align}

\end{document}